\documentclass[a4paper,fleqn]{cas-dc} 

\usepackage[numbers]{natbib}

\def\tsc#1{\csdef{#1}{\textsc{\lowercase{#1}}\xspace}}
\tsc{WGM}
\tsc{QE}
\usepackage{algorithm}
\usepackage{algorithmic}
\usepackage[normalem]{ulem}
\usepackage{graphicx}
\usepackage{subcaption}
\newcommand{\chen}[1]{\textcolor{black}{#1}}

\begin{document}
\let\WriteBookmarks\relax
\def\floatpagepagefraction{1}
\def\textpagefraction{.001}
\let\printorcid\relax    

\shorttitle{CIDGMed}    

\shortauthors{Shunpan L, Xiang L}  

\title [mode = title]{CIDGMed: Causal Inference-Driven Medication Recommendation with Enhanced Dual-Granularity Learning}   

\author[1,2]{Shunpan Liang}
\ead{liangshunpan@ysu.edu.cn}
\credit{Supervision, Funding acquisition}

\author[1,3]{Xiang Li}
\ead{lilixiang222@gmail.com}
\credit{Conceptualization, Methodology, Software, Validation,  Formal analysis, Data Curation, Investigation, Writing – original draft, Writing - Review \& Editing, Visualization}

\author[1]{Shi Mu}
\ead{mushi@stumail.ysu.edu.cn}
\credit{Writing – original draft, Writing - Review \& Editing}

\author[1]{Chen Li}
\ead{lichen36211@gmail.com}
\credit{Supervision, Writing – original draft, Writing - Review \& Editing}
\cortext[1]{Corresponding author.}
\cormark[1]

\author[1]{Yu Lei}
\ead{leiyu0160@gmail.com}
\credit{Supervision}

\author[1]{Yulei Hou}
\ead{ylhou@ysu.edu.cn}
\credit{Resources}

\author[4]{Tengfei Ma}
\ead{tfma@hnu.edu.cn}
\credit{Supervision}

\affiliation[1]{
            organization={Yanshan University},
            city={QinHuangDao},
            postcode={066004}, 
            country={China}
}

\affiliation[2]{
            organization={Xinjiang College of Science \& Technology},
            city={Korla},
            postcode={841000}, 
            country={China}
}

\affiliation[3]{
            organization={Peking University},
            city={Beijing},
            postcode={100091}, 
            country={China}
}


\affiliation[4]{
            organization={Hunan University},
            city={ChangSha},
            postcode={410012}, 
            country={China}
}


\begin{abstract}
Medication recommendation aims to integrate patients' long-term health records to provide accurate and safe medication combinations for specific health states. Existing methods often fail to deeply explore the true causal relationships between diseases/procedures and medications, resulting in biased recommendations. Additionally, in medication representation learning, the relationships between information at different granularities of medications—coarse-grained (medication itself) and fine-grained (molecular level)—are not effectively integrated, leading to biases in representation learning. To address these limitations, we propose the Causal Inference-driven Dual-Granularity Medication Recommendation method (CIDGMed). Our approach leverages causal inference to uncover the relationships between diseases/procedures and medications, thereby enhancing the rationality and interpretability of recommendations. By integrating coarse-grained medication effects with fine-grained molecular structure information, CIDGMed provides a comprehensive representation of medications. Additionally, we employ a bias correction model during the prediction phase to further refine recommendations, ensuring both accuracy and safety. Through extensive experiments, CIDGMed significantly outperforms current state-of-the-art models across multiple metrics, achieving a 2.54\% increase in accuracy, a 3.65\% reduction in side effects, and a 39.42\% improvement in time efficiency. Additionally, we demonstrate the rationale of CIDGMed through a case study.
\end{abstract}

\begin{highlights}
\item We use causal inference to uncover the true causal relationships between diseases/procedures and medications, reducing spurious correlations, thereby minimizing recommendation bias and enhancing the interpretability of recommendations.

\item We consider both coarse-grained information at the entity level and fine-grained information at the molecular level, leveraging the collaboration between this dual-granularity information to enhance the model's learning capacity.

\item We use a bias correction module as a post-processing intervention method during the model recommendation phase to refine the initial results and further optimize the model's predictive performance.

\item Extensive experiments demonstrate that CIDGMed significantly outperforms current state-of-the-art models across multiple metrics, achieving a 2.54\% increase in accuracy, a 3.65\% reduction in side effects, and a 39.42\% improvement in time efficiency.

\item An intricate and elaborate case study based on a real medical record is designed to elucidate the effectiveness and rationality of the model based on causal inference.
\end{highlights}


\maketitle
\section{Introduction}

As society advances and the population grows, the healthcare system is facing unprecedented pressures, including the imbalance of medical resources and the complexity of disease diagnosis and treatment, which have become prominent issues in the medical field. In this context, artificial intelligence (AI)--based medication recommendations~\cite{safedrug, carmen,stratmed} emerge as a key component, bringing new hope to healthcare. Compared to traditional manual methods, this system, based on extensive patient data, can make rapid responses and decisions, significantly addressing the issue of scarce medical resources. Furthermore, by considering patients' individual characteristics, medical history, and current health status~\footnote{Patients' health status encompasses the health status of individuals, including diseases and procedures in healthcare encounters.}, the medication recommendation system can provide accurate and safe medication combinations, playing a key role in alleviating medical pressures and enhancing treatment outcomes.

Compared to recommendation systems (RSs) in other domains, medication recommendation faces unique and significant challenges that prevent the direct application of methodologies from other RSs~\cite{seq2,seq3,session1,session2}. These challenges primarily lie in the following two aspects. Firstly, medication recommendation involves the complex relationships among various medical concepts (medications, diseases, procedures) and the extensive chemical and molecular structure knowledge inherent in medications. Secondly, when evaluating recommendation performance, medication recommendation not only focuses on accuracy but also places a high emphasis on safety concerns. This is because drug-drug~\footnote{The terms "drug" and "medication" are used interchangeably in this paper.} interactions (DDI)s~\cite{ddi1,ddi2,ddi3} can cause side effects in patients.

Early medication RSs~\cite{earlywork1,ealywork2,ealywork3} primarily focused on analyzing patients' current health status, while overlooking their long-term medical histories. Subsequent researches~\cite{dmnc,longitudinal2,longitudinal3} began to recognize the importance of longitudinal medical records and modelled patients' historical records based on time series analysis. Although these studies improved the accuracy of medication recommendations to some extent, they neglected the safety of medication recommendations, which is crucial in this field. Consequently, some studies~\cite{ddi(mtf1),ddi(mtf2)} have considered drug-drug interactions as the main factor affecting safety and have focused on reducing the impact of these interactions to enhance the safety of medication recommendation systems. Despite the significant achievements in existing research, there remain considerable biases in the recommendation results, primarily stemming from the following aspects:

\begin{figure}
    \centering
    \includegraphics[width=0.9\linewidth]{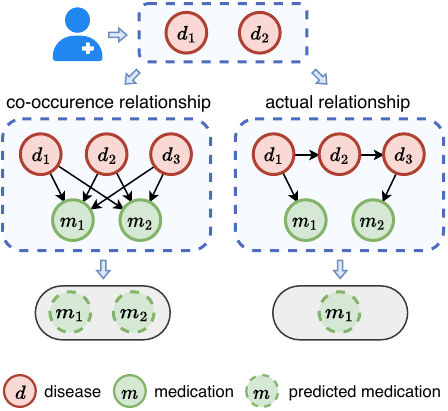}
    \caption{A case where a method based on co-occurrence relationships leads to erroneous recommendation results.}
    \label{fig:intro1}
\end{figure}

(1)  \textbf{Unclear Relationships between Disease/Procedure and Medication.}

Previous co-occurrence-based researches~\cite{co-occur1,co-occur2,aka-safe} assumed that high co-occurrence implies a direct relationship or influence between diseases and medications. However, this is not always the case. As shown in Figure~\ref{fig:intro1}, the results learned using the co-occurrence method indicate a correlation between each disease ($d_1$, $d_2$, $d_3$) and each medication ($m_1$, $m_2$). In reality, the correct relationships should be that $d_1$ causes $d_2$, which in turn causes $d_3$, while $m_1$ specifically targets $d_1$, and $m_2$ specifically targets $d_3$. When models encounter a new combination of diseases ($d_1$, $d_2$), the co-occurrence method mistakenly recommends both $m_1$ and $m_2$, resulting in inaccurate and potentially harmful outcomes.

Hence, these studies~\cite{co-occur1,co-occur2,aka-safe} addressing multi-disease issues relied solely on co-occurrence to determine the relationships between diseases and medications, failing to identify the clear relationships between them. This bias can be amplified through feedback, affecting the accuracy and safety of the recommendations.

(2) \textbf{Unintegrated Dual-Granularity Information.}
In medication recommendation research, there are two main approaches: one is to directly learn the coarse-grained representation of medications from data, and the other, given that most medications consist of multiple molecules~\footnote{\url{https://atcddd.fhi.no/atc/structure\_and\_principles/}}, maps medications to individual molecules and treats them as substitutes for the medication itself, emphasizing the importance of the molecular (fine-grained) level. In particular, using only coarse-grained information \cite{cognet,stratmed} fails to capture the chemical and molecular structure information, making it difficult to identify medications with similar compound structures and to analyze medication-medication interactions. On the other hand, using only fine-grained information \cite{safedrug,molerec}, although it can provide detailed chemical information, the ultimate task of medication recommendation is to recommend the entire medication, not just a single molecule, and this can introduce noise into the training process, affecting the stability and robustness of the model.

However, current methods typically focus only on either the coarse-grained level or the fine-grained level of medications, failing to effectively integrate information from both granularities. This leads to biases in representation learning and recommendation results.

Therefore, to address the above limitations, we design a \textbf{C}ausal \textbf{I}nference-driven \textbf{D}ual-\textbf{G}ranularity \textbf{Med}ication Recommendation method named \textbf{CIDGMed}. The main contributions of our method are summarized as follows:

\begin{itemize}
\item
We introduce causal inference techniques that can uncover the causal relationships between diseases/procedures and medications, enhancing the rationality and interpretability of medication recommendations.
\item
A dual-granularity fusion method is proposed. At the coarse-grained level, it learns the effects of entire medications, while at the fine-grained level, it mines richer relationships from molecular structures.
\item
We employ a post-processing intervention method by introducing a bias correction model. This model further adjusts the generated recommendations using the established causal relationships during the prediction phase, thereby enhancing the accuracy and safety of the model.
\item 
Extensive experiments on real-world datasets demonstrate superior performance compared to existing state-of-the-art models. Our approach achieves a 2.54\% increase in accuracy and a 3.65\% enhancement in safety. Additionally, we demonstrate the interpretability of CIDGMed through a case study. Our code is publicly available at GitHub\footnote{\url{https://github.com/lixiang-222/CIDGMed}}.
\end{itemize}

We provide an overview of the different sections of the paper: (1) Introduction: Introduces the main innovations and the motivation behind this work. (2) Related Work: Summarizes typical studies and current trends in medication recommendations and causal inference in RSs. (3) Problem Definition: Clearly explains the inputs and outputs and the special terms mentioned in this paper. (4) Methods: Presents the core ideas of the model and the specific technical details of its implementation. (5) Experiments: Introduces the experimental background and results, and presents a series of supportive experiments to provide an in-depth analysis of the results. A specific case study of the workflow in this paper can be found in Subsection \ref{subsub:case study}. (6) Conclusion: Summarizes the research findings and offers perspectives on future research directions.

\section{Related Works}
In this section, we describe related work in the areas of medication recommendation and causal inference in RSs.

\subsection{Medication Recommendation}

In recent years, the field of medication recommendation has developed rapidly, and mainstream medication recommendation models can be broadly categorized into three main directions:

The first category mainly focuses on the technical aspect of medication recommendation. The earliest methods~\cite{earlywork1,ealywork2,ealywork3} use statistical data approaches. With the development of collaborative filtering methods, Medicine-LDA~\cite{waghmare2024supervised} proposed a context-aware collaborative model based on Latent Dirichlet Allocation (LDA) to integrate multiple types of information. CompNet~\cite{liang2021compnet} addresses recommendation issues by framing the task as an order-free Markov decision process (MDP) problem. In recent years, deep learning has achieved tremendous success, leading to the emergence of medication recommendation methods using this technology. For example, RETAIN~\cite{retain} is based on a two-level neural attention model that can detect influential past visits and significant information within those visits. The study~\cite{jin2018treatment} develops three different LSTMs to model heterogeneous data interactions for predicting next-period prescriptions. GAMENet~\cite{gamenet} leverages Graph Neural Networks (GNN) to enhance medication recommendation systems. Similarly, COGNet~\cite{cognet} utilizes the Transformer architecture for medication recommendations, adopting a translation approach to infer medications from diseases/procedures.

The second category of research focuses more on the relationships between diseases/procedures and medications. It constructs relational networks to explore the potential connections among various medical entities. For example, DPR~\cite{drug_package} classifies the interactions between different medications and represents them as graphs, designing a medication package recommendation framework based on graph neural networks to integrate medication interaction information. DPG~\cite{drug_package_generation} considers the interaction effects between medications affected by patient conditions, initializes the medication interaction graph based on medical records and domain knowledge, and uses an RNN to capture medication interactions. StratMed~\cite{stratmed} utilizes a dual-property network to address the mutual constraints between the safety and accuracy of medication combinations, synergistically enhancing these two properties.

The third category of research underscores the importance of incorporating domain-specific knowledge to enhance the relationships among various medical entities within the dataset. By integrating more expert knowledge beyond the dataset, these approaches aim to improve the accuracy and safety of medication recommendations. For instance, SafeDrug~\cite{safedrug} introduces detailed molecular information to characterize the relationships between medications. This approach aims to reduce the risk of adverse drug-drug interactions (DDI) and provide safer medication combinations. By focusing on molecular details, SafeDrug enhances our understanding of how different medications interact at a molecular level, contributing to safer prescription practices. Similarly, MoleRec~\cite{molerec} improves the relationships between medications by simulating specific molecular substructures within the medications. This method allows for a more accurate and precise representation of DDIs, enabling more reliable predictions of potential interactions. MoleRec's approach highlights the importance of detailed molecular simulation in understanding medication interactions. And Carmen~\cite{carmen} incorporates patient records into the molecular representation learning process to enhance the ability to differentiate between molecular differences.

Despite the success of the aforementioned methods, they have certain limitations: (1) they primarily rely on the co-occurrence relationships among various medical entities in patient historical records, failing to uncover the underlying causal relationships; (2) they do not consider the use of both coarse-grained medication information and fine-grained molecular information for collaborative supplementary learning and recommendation, which could significantly enhance the accuracy, safety, and interpretability of medication recommendations.

\subsection{Causal Inference in RSs}\label{sec:causal}

In the real world, causality drives the system, prompting researchers in recommender systems to leverage causal inference to extract causal relationships and thereby enhance the RS. Formally, causality can be defined as the relationship between cause and effect, where the cause is partly responsible for the effect ~\cite{wu2022opportunity}. Causal inference, on the other hand, refers to the process of identifying and utilizing causal relationships based on experimental or observational data~\cite{yao2021survey}. Causal relationships are primarily applied in three key stages of RSs:

(1) Data preprocessing: In recent years, some methods~\cite{gao2024causal,zhang2021causal,gao2023cirs} have utilized causal relationships to address issues such as data debiasing or data augmentation.
Specifically, causal theory empowers us to pinpoint the underlying cause of data bias by thoroughly examining the generation process of recommendation data. Hence, through causal inference, some studies can effectively mitigate the impact of bias. (2) Model development: Causal inference contributes to enhancing the interpretability of the model itself~\cite{interpret2,interpret3,si2022model,zheng2022disentangling,missing1,missing2,missing3}. (3) Recommendation outcomes: This is a postprocessing mechanism that incorporates causal relationships into the obtained recommendation results to correct bias issues in the model learning process, thereby enhancing their accuracy and interpretability~\cite{gao2024causal,xian2019reinforcement,tran2021recommending}.

Building on the powerful inferential capabilities of causal inference, this paper innovatively introduces it into the field of medication recommendation. Specifically, we employ a causal inference-based approach to mine potential links between medications and diseases/procedures, eliminating spurious correlations arising from co-occurring relationships and providing more precise causal relationships. Additionally, during the recommendation phase of the model, we leverage the learned causal relationships to correct model biases, thereby enhancing the model's performance and interpretability.

\section{Problem Definition}\label{sec:method}

\subsection{Medication Combination Recommendation}

\textbf{Medical Entity.} Medical entities, in this article, include diseases, procedures, and medications, and these are denoted by the symbols \(\mathcal{D} = \{d_1, d_2, \ldots\}\), \(\mathcal{P} = \{p_1, p_2, \ldots\}\), and \(\mathcal{M} = \{m_1, m_2, \ldots\}\) respectively.

\textbf{Input and Output}. This paper uses Electronic Health Records (EHR) as its data source~\footnote{MIMIC-III and MIMIC-IV are two extensively utilized medical datasets derived from real-world clinical records in intensive care units, employed for the research and analysis of clinical data in critical care settings.}, covering a wide range of patient visits and treatment records. Each patient's record is denoted as \(\mathcal{H}\), containing longitudinal visit records \(\mathcal{H} = \{v_1, v_2, \ldots, v_t\}\), where \(v_t\) is the clinical information associated with the \(t_{th}\) visit. 
Each visit record \(v_t = \{\mathcal{D}_t, \mathcal{P}_t, \mathcal{M}_t\}\) includes three elements: \(\mathcal{D}_t \subset \mathcal{D}\), \(\mathcal{P}_t \subset \mathcal{P}\), \(\mathcal{M}_t \subset \mathcal{M}\), representing the patient's historical and current visit's medical data. If it is the first visit, prior records will be absent. These elements are encoded using a multi-hot encoding technique. Our model's output, denoted as \(\hat{\mathcal{M}}_t\), predicts the medication combination for the visit $v_t$.

\subsection{Safety of Medication Combination}
Safety is particularly important in medication recommendation, as certain medication combinations can pose significant health risks. To ensure the safety of these combinations, it is necessary to incorporate DDI-related knowledge and minimize the occurrence of DDIs as much as possible. Inspired by SafeDrug~\cite{safedrug}, DDI information is extracted from the Adverse Event Reporting System~\cite{AERS}. In our model, DDI information is represented in a matrix format ${\mathbf{M}^{ddi}} \in \{0,1\}^{|\mathcal{M}| \times |\mathcal{M}|}$, where $\mathbf{M}^{ddi}_{m_im_j} = 1$ indicates an interaction between $m_i$ and $m_j$. A higher frequency of DDI indicates a greater potential for safety issues in the recommended results.

\subsection{Causal Discovery and Inference}
As discussed in Section~\ref{sec:causal}, causal discovery and inference are crucial for understanding the interactions between entities~\cite{causal_inference1}, helping to estimate genuine causal effects and thereby enhancing interpretability, particularly in complex data scenarios. Therefore, causal discovery and inference are crucial for identifying the underlying causal relationships between different medical entities. For example, through these uncovered relationships, it can be discovered that Amoxicillin can treat Strep Throat. This not only enhances the accuracy and safety of the recommendations but also explains why Amoxicillin is effective for treating Strep Throat, thereby improving interpretability. Such relationships cannot be well identified through traditional methods that rely on co-occurrence relationships, leading to inappropriate recommendations. For instance, co-occurrence-based methods might recommend Ibuprofen for patients with Strep Throat, resulting in irrational suggestions and introducing biases that impact the performance of the recommendations.
Specifically, we use statistical algorithms and machine learning to exclude meaningless relationships in the data through backdoor~\cite{backdoor} paths, thereby identifying potential causal relationships. Furthermore, causal inference uses observational data to quantify the impacts of these relationships, which is crucial for decision-making and policy formulation. It is important to note that in this paper, the causal mechanism permeates two aspects: the model and the recommendation results. In the model aspect, we leverage a causal graph to explore the relationships among diseases, procedures, and medications, enhancing the model's inferential capabilities. In the recommendation results aspect, we employ learned causal effects to further correct biases in the recommendation results. These specific details will be provided in the following section.

\begin{figure*}[h]
    \centering
    \includegraphics[width=\textwidth]{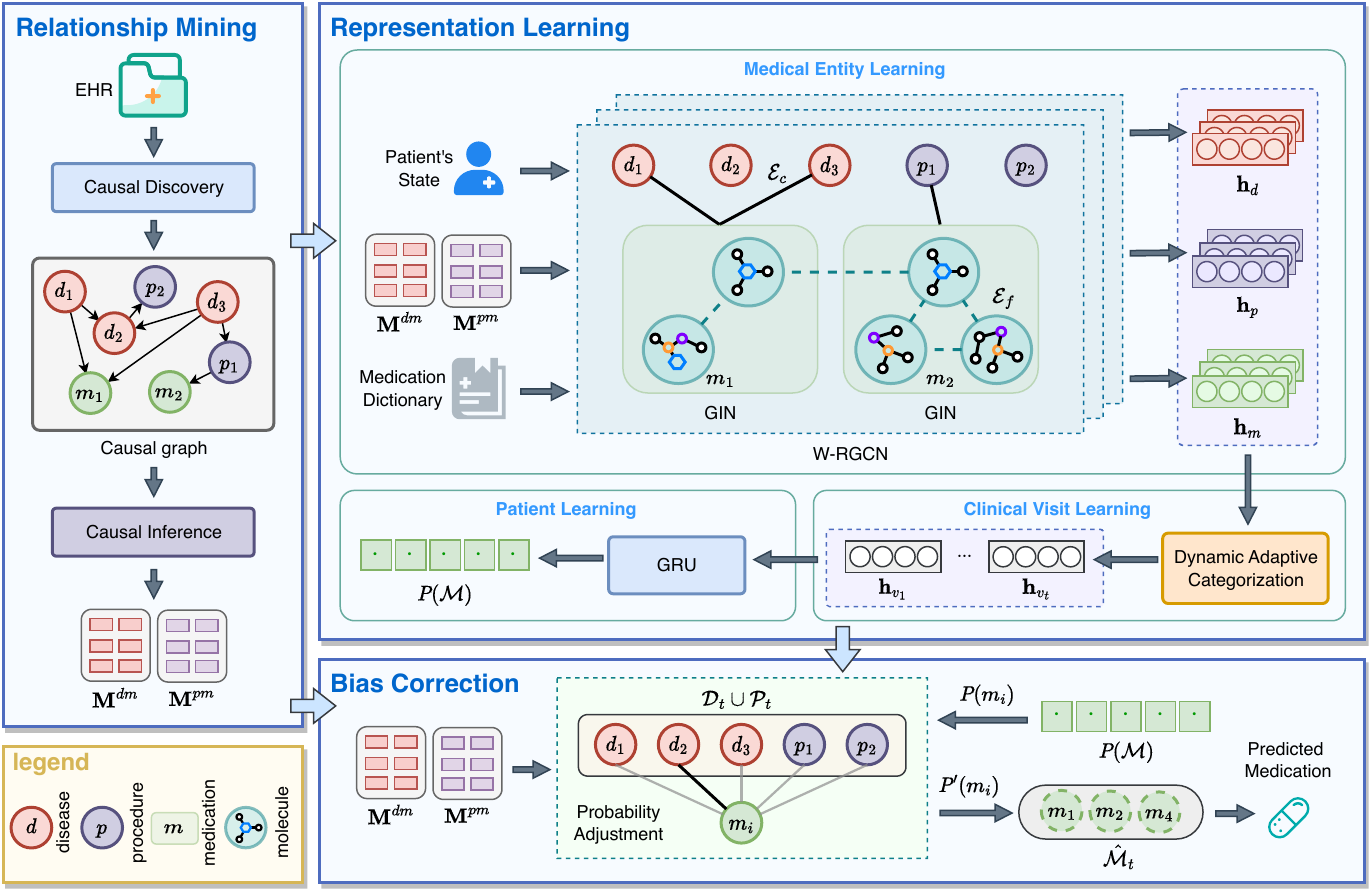}
    \caption{CIDGMed Flowchart: The Relationship Mining module on the left side of the figure utilizes causal discovery and causal inference on the EHR to generate causal graphs and causal effect matrices. The Representation Learning module at the top of the figure builds on this foundation with dual granularity at the medication level and the molecular level to learn patient representations and recommend preliminary medication probabilities. The Bias Correction module at the bottom of the figure corrects the recommended probabilities for each medication based on the causal effects and recommends the final combination of medications.}
    \label{fig:model}
\end{figure*}

\section{Methods}

\begin{table}[h]
\centering
\footnotesize
\caption{Notations and Descriptions.}
\addtolength{\tabcolsep}{1pt}
\begin{tabular}{>{}c|>{}c}
\specialrule{.15em}{.15em}{.15em}
\textbf{Notations}  & \textbf{Descriptions} \\ 
\specialrule{.1em}{.1em}{.1em}
$\mathcal{D},\mathcal{P}, \mathcal{M}$  & disease set, procedure set, medication set\\
$|\mathcal{D}|,|\mathcal{P}|, |\mathcal{M}|$ & number of disease / procedure / medication \\ 
$|\mathcal{S}|$ & number of molecule \\
$d,p,m$ & disease, procedure, medication\\
$\mathcal{H}$ &  patient’s history record\\
$v_t$ & clinical information of the $t$-th visit \\
${\mathbf{M}^{ddi}}$ &  DDI matrix\\ 
$\mathbf{M}^{dm}$, $\mathbf{M}^{pm}$ & causal effect matrice between \\
 &  disease / procedure and medication   \\
 $G$, $G'$ & causal graph, enhanced graph\\
 $X_i$ & the $i$-th entity in Bayesian network\\
 $\mathbf{h}_{d_i}, \mathbf{h}_{p_j}$ & embedding of entity $d_i$ / $p_j$ \\
 $\mathcal{R}$ & relevance stratification structure \\
 $r^{dm}_l, r^{pm}_l$  &  relationship between disease \\ 
 &  / procedure and medication at the $l$-th layer \\
 $\mathbf{A}$ & medication-molecule relationship matrix \\
  $a_{ij}$ & importance of molecule$s_j$ in medication $m_i$.\\
  $\mathbf{h}_{s_j}$ & molecule embedding of $s_j$\\
  $\mathbf{h}_{m_i}$ & medication embedding of $m_i$ \\
  $\mathcal{N}^{en}$ & set of medical entities\\
  $\mathcal{N}^{mo}$ & set of molecules\\
  $\mathcal{E}^c$ & relationships between medical entities\\
  $\mathcal{E}^f$ & relationships between
  molecules\\
 $\mathbf{h}^{mo,{l}}_i$ & representation of molecule $i$ at layer $l$\\
 $\mathbf{h}^{en,{l}}_i$ & representation of entity $i$ at layer $l$ \\
 $\mathbf{h}_{v_t}$ & final represen-
tation of the current visit $v_t$\\
 $\mathbf{h}_{H}$ & patient representation \\
 $\tau_1, \tau_2$ & magnitude of adjustment required\\
 $\delta_1, \delta_2$ & threshold value of causal matrix\\
 $P(m_i)$ & recommendation probability of $m_i$ \\
 $P'(m_i)$ & recommendation probability of $m_i$ \\
 & after bias correction\\
\specialrule{.15em}{.15em}{.15em}
\end{tabular}\label{tab:notations}
\end{table}

Figure~\ref{fig:model} illustrates our model, which consists of three main components. In this study, we aim to mine and utilize the causal relationships among various medical entities across three distinct stages to achieve personalized medication recommendations. For ease of reading, we list important notations and their explanations in Table~\ref{tab:notations} of this paper.

First, in the \textbf{Relationship Mining} stage, we extract patient information (i.e., diseases, procedures, and medications) from EHRs. Using causal discovery, we learn the causal relationships among these medical entities and construct personalized causal graphs for each patient. This process involves obtaining the causal effect matrices $\mathbf{M}^{dm}$ and $\mathbf{M}^{pm}$, which represent the relationships between diseases/procedures and medications. Next, in the \textbf{Representation Learning} stage, we refer to a medication dictionary \footnote{\url{https://go.drugbank.com/releases/latest}} to establish mappings between each medication and its molecular components. 
We employ the causal effect matrix constructed during the relationship mining phase (based on the coarse-grained relationships at the medication level) and integrate the relationships between molecules and medications, as well as between molecules themselves (based on the fine-grained relationships at the molecular level), to form a dual-granularity representation message passing method.
These representations help the study categorize and aggregate medical entities within the causal graph more precisely, forming a visit-level detailed representation $\textbf{h}_v$. By incorporating patients' historical data, we ultimately derive the medication recommendation probability $P(\mathcal{M})$. Finally, in the \textbf{Bias Correction} stage, we adjust the probabilities for each medication based on the initial causal matrices $\mathbf{M}^{dm}$ and $\mathbf{M}^{pm}$ to correct biases that arise during the model training process, thus recommending medication combinations that are both effective and safe.

\subsection{Relationship Mining}

At this stage, our objective is to leverage causal discovery to deeply investigate the relationships between medical entities and to quantify these causal relationships as causal effects through causal inference. This will lay the essential groundwork for graph networks in subsequent representation learning and for bias correction methods in the recommendation phase.

Firstly, we explore the relationships among homogeneous entities, that is, entities of the same type, such as disease $d_2$ being caused by disease $d_1$. Based on the data distribution $U$ within the EHR, we employ a causal discovery algorithm to construct causal graphs among the same type of medical entities. Given the extensive and discrete nature of medical data, we select the Greedy Intervention Equivalence Search (GIES) algorithm~\cite{GIES} for the causal discovery module. This method not only learns from the data but also iteratively optimizes Bayesian equivalence classes~\cite{bayesian}, enabling us to accurately determine the most suitable causal graph. Finally, we use an equivalence score criterion $\mathcal{F}(G, U)$, which generates Bayesian equivalence classes, to assess the quality of the generated causal graph.

\begin{algorithm}
\caption{The Algorithm of Causal Discovery}
\begin{algorithmic}[1]
\STATE \textbf{Input}: data distribution of EHR $U$
\STATE \textbf{Output}: causal graph $G'$
\STATE randomly initialize a graph $G$
\STATE \textbf{do}:
\STATE \hspace{0.5cm} $\mathcal{F}(G, U)$ $\leftarrow$ compute equivalent classes (eq\ref{eq:bayes})
\STATE \hspace{0.5cm} $G'$ $\leftarrow$ update causal graph (eq\ref{eq:gies})
\STATE \textbf{while} $G' \neq G$
\STATE \textbf{return} causal graph $G'$
\end{algorithmic}
\label{alg:gies}
\end{algorithm}

\begin{figure}
    \centering
    \includegraphics[width=0.8\linewidth]{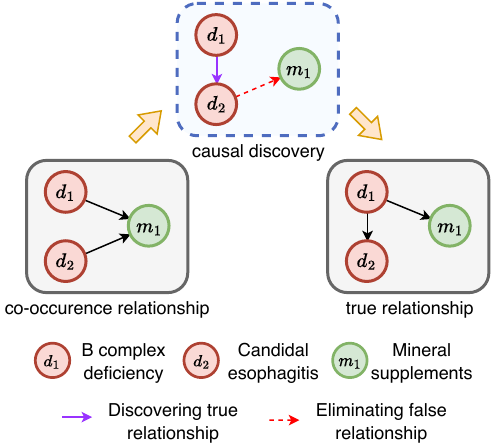}
    \caption{A real example of using causal discovery to correct incorrect relationships from co-occurrence relationships to generate true relationships.}
    \label{fig_causal}
\end{figure}

As depicted in Figure~\ref{fig_causal}, traditional co-occurrence-based methods suggest that both $d_1$ and $d_2$ are associated with $m_1$. However, through causal discovery, we determine that $d_2$ is a disease caused by $d_1$, and $m_1$ is a medication specifically prescribed for $d_1$, with no direct relevance to $d_2$. The causal relationships established by this approach, based on the backdoor criterion~\cite{backdoor}, effectively eliminate spurious associations between medications and diseases.
The entire causal discovery process is shown in algorithm \ref{alg:gies}.

These findings provide a robust foundation for the detailed examination of the relationships between medications and diseases or procedures in future tasks. The analytical approach applied is captured in the following formula:
\begin{gather}
    \label{eq:bayes}
    \mathcal{F}(G, U) = \sum_{i=1}^{n} f(X_i, \mathcal{Q}^G_{X_i}),\\
    \label{eq:gies}
    G' = \text{GIES}(\mathcal{F},G),
\end{gather}
where \(n\) represents the number of medical entities (i.e., diseases, procedures, and medications) in the Bayesian network, \(X_i\) denotes the \(i\)-th entity in the network, and \(\mathcal{Q}^G_{X_i}\) stands for the parents of \(X_i\) in the graph \(G\), representing the set of entities that directly influence \(X_i\) according to the Bayesian network's structure. $f(\cdot)$ is a Bayesian evaluation function that evaluates the relationship between a given node and its parent node in a Bayesian network. The \(\text{GIES}(\cdot)\) optimizes and learns an enhanced graph \(G'\) from the initial graph \(G\) and the equivalence score \(\mathcal{F}\).

In addition to causal relationships existing among the same medical entities, there are also causal relationships between different medicinal entities. As shown in Figure~\ref{fig_causal}, $m_1$ is a medication specifically for $d_1$. Hence, for heterogeneous relationships, we utilize causal inference to demonstrate the impact of medications on diseases or procedures. This straightforward approach clearly shows whether medications are effective against various diseases or procedures, thereby uncovering their direct relationships. Specifically, the causal graph $G'$ is represented as a binary variable, and a discretized Generalized Linear Model (GLM) \cite{glm} is employed to model the causal effects between diseases/procedures and medications. The specific formula is as follows:
\begin{equation}
    \label{eq:glm}
    \text{GLM}(\mu) = \beta_0 + \beta_1 X_1 + \beta_2 X_2 + \ldots + \beta_n X_n,
\end{equation}
where \(\text{GLM}(\cdot)\) denotes the logit link function, \(\mu\) represents the probability of a binary variable, \(\beta_0\) is the intercept, and \(\beta_1, \beta_2, \ldots, \beta_n\) are the corresponding coefficients. \(X_1, X_2, \ldots, X_n\) represent the model's independent variables, including the records of diseases/procedures and medications. In this model, the probability \(\mu\) of the binary variable indicates the likelihood of a positive response from medications to patients with identified diseases or procedures, elucidating the causal effect of a specific disease or procedure on medication. Ultimately, the causal effect matrices for diseases-medications and procedures-medications can be obtained, denoted as \(\mathbf{M}^{dm} \in \mathbb{R}^{|\mathcal{D}| \times |\mathcal{M}|}\) and \(\mathbf{M}^{pm} \in \mathbb{R}^{|\mathcal{P}| \times |\mathcal{M}|}\), respectively.

\subsection{Representation Learning}
The main objective of this phase is to obtain detailed entity representations, visit representations, and ultimately comprehensive patient representations, built upon multiple causal-based relational networks, and to recommend preliminary medication probabilities.

\subsubsection{Medical Entity Learning}
This subsection describes the construction of dual-granularity relationship network and the process of obtaining entity representations.

In the real-world diagnosis and treatment process, doctors determine prescriptions based on the patient's current health status and their historical medical records. In our model, we use the current disease \(\mathcal{D}_t\) and procedure \(\mathcal{P}_t\) to represent the current health status and regard the diseases \(\mathcal{D}_{t-1}\), procedures \(\mathcal{P}_{t-1}\) and medications \(\mathcal{M}_{t-1}\) from the last visit as the historical medical record. These combined data sources provide a comprehensive overview of the current visit, helping us to simulate the real-world medical consultation process. We initially establish embedding tables \(\mathbf{E}_d \in \mathbb{R}^{|\mathcal{D}| \times dim}\) and \(\mathbf{E}_p \in \mathbb{R}^{|\mathcal{P}| \times dim}\) for diseases and procedures, with each row corresponding to a specific disease or procedure.
\begin{equation}
    \label{eq:entityrepr_d/p}
    \mathbf{h}_{d_i} = \mathbf{E}_d(d_i), \quad 
    \mathbf{h}_{p_j} = \mathbf{E}_p(p_j), 
\end{equation} 
where \(d_i \subset \mathcal{D}_t\), \(p_j \subset \mathcal{P}_t\) represent specific medical entities, \(\mathbf{h}_{d_i} \in \mathbb{R}^{dim}\), \(\mathbf{h}_{p_j} \in \mathbb{R}^{dim}\) represent the embedding for entity \(d_i\) and \(p_j\), respectively.

\textbf{Coarse Granularity}:
First, we establish and apply a coarse-grained relationship network. Based on the previously generated \(\mathbf{M}^{dm}\) and \(\mathbf{M}^{pm}\), we introduce a relevance stratification strategy \cite{stratmed}. This strategy adaptively layers all relationships in \(\mathbf{M}^{dm}\) and \(\mathbf{M}^{pm}\) according to their values, using a gradient \(\mathcal{K}\) to generate an \(n\)-layered relevance stratification structure \(\mathcal{R}^{dm}=\{r^{dm}_1, r^{dm}_2, \ldots, r^{dm}_n\}\) and \(\mathcal{R}^{pm}=\{r^{pm}_1, r^{pm}_2, \ldots, r^{pm}_n\}\). The stratification follows a pyramid structure, with more relationships at the lower layers and fewer at the upper layers. By assigning different weights to each layer in subsequent steps, we enhance the learning strength for rare data, addressing the imbalance in the data and obtaining coarse-grained relationships. The specific formulas are as follows:
\begin{equation}
    |r^{dm}_j| = |r^{dm}_1| \mathcal{K}^{j-1}, \quad |r^{pm}_l| = |r^{pm}_1| \mathcal{K}^{l-1},
\end{equation}
where, \(r^{dm}_l\) refers to the \(l^{th}\) layer of the medication-disease relationship, \(|r^{pm}_l|\) denotes the number of relationships at this layer, \(\mathcal{K}\) represents the hierarchical gradient. 

\textbf{Fine Granularity}:
Next, we mine fine-grained information, i.e., detailed information originating from the molecular structure level. By importing a pre-defined dictionary of medications, we obtain a precise mapping between medications and their molecular structures. Different structures may exist within the same medication, indicating that similar actions may be triggered; hence there is some connection between the molecular structures of the medication. 

At the same time, most medications are composed of multiple molecular structures, and the same molecular structure may also exist in different medications, sharing the same representation. Therefore, we develop a learnable medication-molecule relationship matrix (\(\mathbf{A} \in \mathbb{R}^{|\mathcal{M}| \times |\mathcal{S}|}\)), where \(a_{ij} \subset \mathbf{A}\) represents the importance of molecule \(s_j\) in medication \(m_i\).

We apply a method similar to that used for constructing entity embeddings to build a molecule embedding table \(\mathbf{E}_s \in \mathbb{R}^{|\mathcal{S}| \times dim}\). And, to accurately represent the medications, we construct medication embeddings (\(\mathbf{h}_{m}\)) based on their molecular composition structures and the importance of each molecule, thereby constructing fine-grained connections between medications at the structural level.
\begin{gather} 
    \label{eq:entityrepr_s}
    \mathbf{h}_{s_j} = \mathbf{E}_s(s_j),\\
    \label{eq:entityrepr_m}
    \mathbf{h}_{m_i} = \sum^{|\mathcal{S}|}_{j=1} a_{ij} \cdot \mathbf{h}_{s_j},
\end{gather}
where \(\mathbf{h}_{s_j} \in \mathbb{R}^{\text{dim}}\) represents the molecule \(s_j\), \(\mathbf{h}_{m_i}\) is the embedding of medication \(m_i\), \(a_{ij}\) represents learnable weight.

\textbf{Dual Granularity Fusion}:
To integrate relationships of different granularities, we use \(\mathcal{G} = (\mathcal{N}^{en}, \mathcal{N}^{mo}, \mathcal{E}^c, \mathcal{E}^f)\), a bipartite graph containing both coarse-grained and fine-grained information. \(\mathcal{G'}\) is a bipartite graph where \(\mathcal{N}^{en}\) represents the set of all medical entities (coarse-grained medications) contained in the visit records, \(\mathcal{N}^{mo}\) represents the set of molecules (fine-grained medications) involved in the medications from the visit records, \(\mathcal{E}^c\) represents the relationships between medical entities (coarse-grained relationships), and \(\mathcal{E}^f\) represents the relationships between molecules (fine-grained relationships).

This graph starts learning at the fine-grained relationship level, and medications are used as intermediaries, interacting with information from other medical entities through coarse-grained relationships, thereby enriching the representation of medications at different granularities. Specifically:

First, for the fine-grained relationships, we construct fully connected graphs between molecular structures belonging to the same medication and use Graph Isomorphism Networks (GIN) \cite{gin} for efficient message passing. This approach updates the representation  $\textbf{h}^{mo} = \{\textbf{h}_s\}$ of the molecule within the medication, significantly enhancing the fine-grained relationships between the molecules. The specific formula is as follows:
\begin{equation}
    \label{eq:propagation_finer}
    \mathbf{h}^{mo,{l+1}}_i = \sigma \left( (1+\epsilon)
    \mathbf{h}^{mo,l}_i + \sum_{j \in \mathcal{N}^{mo}_{e^f,i}} \mathbf{h}^{mo,l}_j \right), \\
\end{equation}
where \(i,j \in \mathcal{N}^{mo}\) is a node in the node set \(\mathcal{N}^{mo}\), \(\mathbf{h}^{mo,{l+1}}_i\) represents the updated feature representation of molecule node \(i\) at layer \(l+1\), \(\epsilon\) is a learnable parameter that controls the influence of the original feature representation, \( \mathcal{N}^{mo}_{e^f,i}\) denotes the set of neighbouring molecules of \(i\) within the medication, \(\mathbf{h}^{mo,l}_i\) represents the original feature representation of molecule \(i\) at layer \(l\), and \(\mathbf{h}^{mo,l}_j\) represents the original feature representation of neighboring molecule \(j\) at layer \(l\).

Next, for the coarse-grained relationships, we perform adaptive adjustments based on the Relational Graph Convolutional Network \cite{RGCN}, mapping the \( n \) levels of entity relevance \( \mathcal{R}^{pm} \) and \( \mathcal{R}^{dm} \) onto the \( \mathcal{E}^c \) of graph \( \mathcal{G'} \). By coarse-grained partitioning of the same relationships, we construct a Weighted-RGCN (W-RGCN) with coarse-grained weights. Through message passing on \( \mathcal{G'} \), we update the representations of diseases, procedures, as well as the representations of medications $\textbf{h}^{en} = \{\textbf{h}_d, \textbf{h}_p, \textbf{h}_m\}$. In the face of coarse-grained relationships, the specific formula is as follows:
\begin{gather}
    \label{eq:propagation_coarse}
    \mathbf{h}_i^{en,(l+1)} = \sigma \left( \sum_{e^c \in \mathcal{E}^c} \frac{1}{q^{en}_{e^c,i}} \sum_{j \in \mathcal{N}^{en}_{e^c,i}} \mathbf{W}_{e^c}^l \mathbf{h}_j^{en,l} \right), \\
    \mathbf{W}_{e^c}^l = r_{e^c} + \Delta \mathbf{W}_{e^c}^l,
\end{gather}

where \(i,j \in \mathcal{N}^{en}\) is a medical entity in the entity node set \(\mathcal{N}^{en}\), \( \mathbf{h}_i^{en,(l+1)} \) is the updated representation of the node \( i\) at layer \( l+1 \), \( \sigma \) is the activation function, \( \mathcal{E}^c \) is the set of edges related to the medical entities, \( q^{en}_{e^c,i} \) is a learnable normalization constant, ensuring the output features are appropriately scaled, \( \mathcal{N}^{en}_{e^c,i} \) is the set of neighboring nodes of \( i \) related to edge \( e^c \), \( \mathbf{W}_{e^c}^l \) is the weight matrix for edge \( e^c \) at layer \( l \), and \( r_{e^c} \) and \( \Delta \mathbf{W}_{e^c}^l \) represents the relationship relevance and the weight update matrix for edge \( e^c \) at layer \( l \), respectively.

Finally, within graph $G'$, we will propagate the molecular representations obtained from fine-grained relational learning to medications and further propagate the learning on coarse-grained relationships. This fusion of coarse and fine-grained propagation is fed back into the network in a recursive loop, achieving a dual-granularity fusion. Ultimately, we obtain the representations of various medical entities (diseases, procedures, and medications) integrating both granularities, respectively: $\textbf{h}_d, \textbf{h}_p$, and $ \textbf{h}_m$.

\subsubsection{Clinical Visit Learning}

Due to the varying physical constitutions of different patients, the same disease can have different impacts on different individuals. Moreover, different diseases also influence each other, and this varies from person to person. Based on this, the study employs a Dynamic Adaptive Categorization mechanism (DAC)~\cite{causalmed} to learn about the effects of diseases in various clinical contexts, thereby enhancing personalized learning for patients.

\begin{figure}
    \centering
    \includegraphics[width=0.8\linewidth]{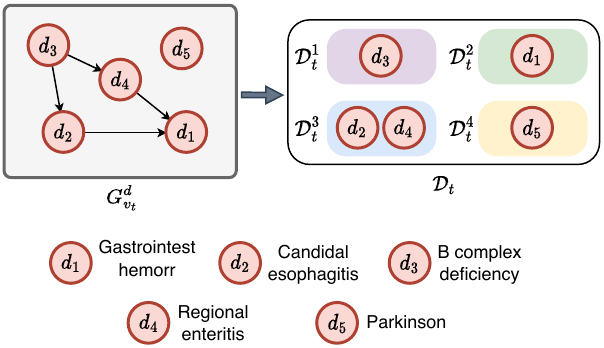}
    \caption{A real example of DAC, we analyze the causal position of each node in the causal graph \( G^d_{v_t} \) to gauge its influence on the patient's current state. For instance, if \( d_3 \) is identified as a causal disease, it is classified as \( {D}^1_t \).}
    \label{fig_dac}
\end{figure}

DAC learns interactions between similar medical entities from causal graphs and integrates them into clinical visit representations. As shown in Figure \ref{fig_dac}, taking diseases as an example, based on the causal positions of entities in the session graph \(G^d_{v_t}\), we divide \(\mathcal{D}_t\) into four sets $\mathcal{D}^j_t$: (1) Causal Disease $\mathcal{D}^1_t$: Diseases capable of causing the occurrence of other diseases, acting as the root cause in the pathway, representing the primary diseases in this clinical visit. (2) Effect Disease $\mathcal{D}^2_t$: Diseases influenced by other diseases, serving as the result in the pathway, representing secondary symptoms in the visit. (3) Middle Disease $\mathcal{D}^3_t$: Diseases capable of causing the occurrence of other diseases while being influenced by other diseases, acting as intermediate nodes in the causal pathway. (4) Independent Disease $\mathcal{D}^4_t$: Diseases existing independently in this visit without a direct causal relationship with other diseases.

Based on DAC, entities can be categorized into four types according to their positions in the causal graph of each user. This paper uses diseases as an example, with medications and treatments being similar. As shown in Figure~\ref{fig_dac}, the four categories of diseases are as follows: (1) Source Diseases ($\mathcal{D}^1_t$), which trigger other diseases and are the cause of some diseases; (2) Intermediate Diseases ($\mathcal{D}^3_t$), which can be caused by other diseases and can also cause other diseases; (3) Resultant Diseases ($\mathcal{D}^2_t$), which can be caused by other diseases; (4) Independent Diseases ($\mathcal{D}^4_t$), which have no causal relationships with other diseases. The specific formula is as follows:

\begin{equation}
    \mathcal{D}^j_t = \text{Classify}(d_i, G^d_{v_t}),
\end{equation}
where \(\text{Classify}(\cdot)\) assigns \(d_i\) to a specific category \(\mathcal{D}^j_t\).

Since the same disease occupies different positions in the causal graphs of different patients, resulting in different categorizations by DAC, they are classified into different categories. Therefore, this paper enhances user personalization by dynamically adjusting the impact of diseases on patients and assigning different weights to various disease categories. The specific formula is as follows:
\begin{gather}
    w^j_t = \frac{\exp(\mathbf{W} \cdot \textbf{h}_{\mathcal{D}^j_t} + b)}{\sum_{k=1}^{4} \exp(\mathbf{W} \cdot \textbf{h}_{\mathcal{D}^k_t} + b)},\\
    \label{eq:aggregate}
    \textbf{h}_{\mathcal{D}_t} = \sum_{i=1}^{|\mathcal{D}_t|}\textbf{h}_{d_i} \cdot w^j_t,
\end{gather}

where $w^j_t$ is the adaptive weight, $\mathbf{W}$ and $\mathbf{b}$ are the trainable weight matrix and bias term, and $\mathbf{{h}_{\mathcal{D}_t}}$ is the final representation of the diseases. Similarly, the final representations for procedures and medications can also be obtained, denoted as $\mathbf{{h}_{\mathcal{P}_t}}$ and $\mathbf{{h}_{\mathcal{M}_t}}$ respectively. Therefore, the final representation of the visit is as follows:

\begin{equation}
    \textbf{h}_{v_t} = [\textbf{h}_{\mathcal{D}_t}||\textbf{h}_{\mathcal{P}_t}||\textbf{h}_{\mathcal{M}_t}].
\end{equation}

\subsubsection{Patient Learning}
Given the robust capability of Gated Recurrent Unit (GRU) \cite{gru} to manage complex sequential challenges, this study employs GRUs to analyze historical records, aiming to elucidate the sequential dependencies inherent in the progression of patients' historical diseases, procedures and medications $\{\mathbf{h}_{v_1},\mathbf{h}_{v_2},\ldots,\mathbf{h}_{v_t}\}$. Then, a Multi-Layer Perceptron (MLP) with powerful nonlinear modelling capabilities is used to obtain the final representation of the patient. The specific formula is as follows:
\begin{gather}
    \mathbf{o}_{v_t} = \text{GRU}(\mathbf{o}_{v_{t-1}},\mathbf{h}_{v_t}),\\
    \mathbf{h}_{H} = \text{MLP}(o_{v_t})
\end{gather}
where \(\mathbf{o}_{v_{t-1}}\) is an intermediate variable generated by the $\text{GRU}(\cdot)$, and we represent \(\mathbf{o}_{v_0}\) with a zero vector. Finally, we use a nonlinear activation function $\sigma$ to transform the patient embeddings to the predicted probability $P(\mathcal{M})$ for each medication $m_i$.
\begin{equation}
    P(m_i)=\sigma(\mathbf{h}_{H}),
\end{equation}
where $\mathbf{h}_{H}$ is the patient representation, and $\sigma$ is a nonlinear activation function that is used to obtain the predicted probabilities for the medication $m_i$.

\subsection{Bias Correction}
Due to the inevitable errors that may arise during the model training process, to mitigate these errors, this paper applies bias correction to the predicted medications using the previously generated causal relationships, thereby more accurately capturing the impact of medications on specific diseases or procedures. Specifically, based on the causal effect matrices $\mathbf{M}^{dm}$ and $\mathbf{M}^{pm}$, for any medication $m_i$ that has a significant therapeutic effect on a specific disease or procedure, we encourage its recommendation and increase its recommendation probability. Conversely, if the medication's therapeutic effect on any disease or procedure during this visit is below a certain quantified treatment effect threshold, we prefer to decrease its probability.
\begin{equation}
    \label{eq:bias_correct}
    P'(m_i) = 
    \begin{cases}
        P(m_i)+\tau_1 ,& max\{\mathbf{M}^{dm}_{m_i-D_t},\mathbf{M}^{pm}_{m_i-P_t}\}\ge \delta_1  \\
        P(m_i)        ,& \delta_1 > max\{\mathbf{M}^{dm}_{m_i-D_t},\mathbf{M}^{pm}_{m_i-P_t}\} \ge \delta_2 \\
        P(m_i)-\tau_2 ,& max\{\mathbf{M}^{dm}_{m_i-D_t},\mathbf{M}^{pm}_{m_i-P_t}\} < \delta_2 
    \end{cases},
\end{equation}
where $\tau_1$ and $\tau_2$ represent the magnitudes of adjustment required for the predicted probability of a medication, while $\delta_1$ and $\delta_2$ are the thresholds derived from the causal matrix. When these thresholds are met, it indicates that the medication has a beneficial effect on the disease (or procedure), thus warranting a correction to the original predicted probability $P(m_i)$ of the medication. The adjusted probability, $P'(m_i)$, is the recommended probability after applying bias correction.
Finally, based on the recommended probability of each medication, we generate the final medication combination.

\subsection{Model Training}
We define the medication recommendation process as a multi-label binary classification task and employ both the binary cross-entropy loss function \(L_{bce}\) and the multi-label margin loss function \(L_{multi}\). Additionally, a DDI loss, \(L_{ddi}\), is implemented to enhance the safety of medication recommendations by calculating the occurrence probability of medication pairs with potential DDI risk within the medication combination. The specific formulas for the three loss functions are as follows:
\begin{gather}
    \mathcal{L}_{bce} = -\sum_{i=1}^{|\mathcal{M}|}{m_i}\log({\hat{m}_i})+(1-{m_i})\log (1-{\hat{m}_i}),\\
    \mathcal{L}_{multi} = \sum_{i,j:{m_i}=1,{m_j}=0} \frac{\max(0,1-({\hat{m}_i}-{\hat{m}_j}))}{|\mathcal{M}|},\\
    \mathcal{L}_{ddi} = \sum_{i=1}^{|\mathcal{M}|}\sum_{j=1}^{|\mathcal{M}|} \mathbf{M}^{ddi}_{m_im_j}\cdot{\hat{m}_i}\cdot {\hat{m}_j},
\end{gather}    
where $m_i$ represents the ground truth, and $\hat{m}_i$ represents the model's predicted value for the medication.

A lower DDI can enhance the safety of recommendations, but pursuing the lowest DDI rate without considering the clinical context may compromise the effectiveness of the prescription. In other words, further blindly reducing DDI would decrease the efficacy of the medication combination, and there is no need to reduce DDI further. Therefore, in constructing the loss function, methods consistent with prior research~\cite{molerec} are employed to ensure a balanced consideration of each loss function.
\begin{gather}
    \label{eq:loss}
    \mathcal{L} = \alpha (\beta \mathcal{L}_{bce}+(1-\beta)\mathcal{L}_{multi})+(1-\alpha )\mathcal{L}_{ddi},\\
    \alpha =     
    \begin{cases}
        1 ,& \mathbf{rate}_{ddi}\le \gamma   \\
        \max\{0, 1- \frac{\mathbf{rate}_{ddi}-\gamma}{kp}\} ,& \mathbf{rate}_{ddi}> \gamma
    \end{cases},
\end{gather}  
where \(\beta\) represents hyperparameters that control model complexity or regularization strength, \(\alpha\), related to \({rate}_{ddi}\), serves as a controllable factor that adjusts the influence of the DDI rate on the model's decision-making process. \(\gamma\), ranging from 0 to 1, signifies a DDI acceptance rate that determines the threshold at which the predicted DDI level is deemed acceptable for clinical use. Lastly, \(kp\) acts as a correction factor, adjusting the strength of the information based on proportionality. 

The overall optimization process of the algorithm is shown in Algorithm \ref{alg:cidgmed}.

\begin{algorithm}
\caption{The Algorithm of CIDGMed}
\begin{algorithmic}[1]
\STATE \textbf{Input}: patient data $\mathcal{H}$, training epoch $ep$
\STATE \textbf{Output}: predicted value of medication $\hat{m}$
\STATE randomly initialize all model parameters
\STATE preprocess the EHR data
\STATE
\STATE \textit{\#relationship mining}
\STATE $G'$ $\leftarrow$ GIES generate causal graph (eq\ref{eq:gies})
\STATE $\mathbf{M}$ $\leftarrow$ GLM generate casual effect (eq\ref{eq:glm})
\STATE
\STATE \textbf{for} $i$ \textbf{in} $ep$:
\STATE \hspace{0.5cm} \textit{\#medical entity learning}
\STATE \hspace{0.5cm} $\mathbf{h}_{d}$ $\leftarrow$ compute disease representation (eq\ref{eq:entityrepr_d/p})
\STATE \hspace{0.5cm} $\mathbf{h}_{p}$ $\leftarrow$ compute procedure representation (eq\ref{eq:entityrepr_d/p})
\STATE \hspace{0.5cm} $\mathbf{h}_{s}$ $\leftarrow$ compute molecular representation (eq\ref{eq:entityrepr_s})
\STATE \hspace{0.5cm}  $\mathbf{h}_{m}$ $\leftarrow$ aggregate $\mathbf{h}_{s}$ into medication (eq\ref{eq:entityrepr_m})
\STATE
\STATE \hspace{0.5cm} \textit{\#fine granularity fusion}
\STATE \hspace{0.5cm} pass messages between $\mathbf{m}$(eq\ref{eq:propagation_finer})
\STATE \hspace{0.5cm} \textit{\#coarse granularity fusion}
\STATE \hspace{0.5cm} pass messages between $\mathbf{d}$/$\mathbf{p}$ and $\mathbf{m}$ (eq\ref{eq:propagation_coarse})
\STATE
\STATE \hspace{0.5cm} \textit{\#clinical visit learning}
\STATE \hspace{0.5cm} $h_\mathcal{H}$ $\leftarrow$ integrate entity via causal graph $G'$ (eq\ref{eq:aggregate})
\STATE
\STATE \hspace{0.5cm} \textit{\#bias correction}
\STATE \hspace{0.5cm} $\hat{m}$ $\leftarrow$ correct bias using causal effect $\mathbf{M}$ (eq\ref{eq:bias_correct})
\STATE
\STATE \hspace{0.5cm}update model parameters with loss (eq\ref{eq:loss})
\STATE \textbf{return} predicted value of medication $\hat{m}$
\end{algorithmic}
\label{alg:cidgmed}
\end{algorithm}

\section{Experiments}
In this section, we conduct extensive experiments on our proposed CIDGMed, aiming to answer the following research questions (RQ):

\begin{itemize}
    \item \textbf{RQ1}: Does CIDGMed significantly outperform other state-of-the-art models in terms of accuracy, safety, and time efficiency?
    
    \item \textbf{RQ2}: Do the key modules proposed in this paper enhance the model's performance?
    
    \item \textbf{RQ3}: Can the dual-granularity architecture in our proposed model recommend medications better than a single-granularity architecture?

    \item \textbf{RQ4}: What specific structures are captured by our causal inference method, and how do they contribute to the model's performance?

    \item \textbf{RQ5}: Why does our approach enhance both effectiveness and interpretability simultaneously?
\end{itemize}

\subsection{Datasets}

\begin{table}
    \caption{Statistics of the datasets.}
    \begin{tabular}{|c|c|c|}
    \toprule
    Item & MIMIC-III & MIMIC-IV \\
    \midrule
    \# patients         & 6,350  & 60,125 \\
    \# visits  & 15,032 & 156,810 \\
    \# diseases         & 1,958  & 2,000 \\
    \# procedures       & 1,430  & 1,500 \\
    \# medications      & 131   & 131\\
    avg. \# of visits    & 2.37  & 2.61\\
    avg. \# of medications & 11.44  & 6.66\\
    \bottomrule
    \end{tabular}
    \label{tab:datasets}
\end{table}

As shown in Table~\ref{tab:datasets}, this paper utilizes the MIMIC-III~\cite{mimic3} and MIMIC-IV~\cite{mimic4} datasets, which are widely used in clinical research and analysis in the Intensive Care Unit (ICU). The data include comprehensive ICU patient clinical records, physiological monitoring data, laboratory test results, and medication records, among other information. We adopt the same data preprocessing method as described in previous work~\cite{causalmed}. For the MIMIC-III dataset, we utilize ICD-9 codes for disease and procedure records, while for the MIMIC-IV dataset, we use both ICD-9 and ICD-10 codes for the relevant data, with medications mapped to ATC-3 codes. We retain only those visit records that include information on diseases, procedures, and medications.

\subsection{Baselines}
To validate our model, we select the following high-performing methods as baseline models for comparison.

\textbf{LR} \cite{lr} (Logistic Regression) is a linear classification algorithm that estimates the probability of an outcome belonging to a certain category by a linear combination of input features, widely used in probability prediction and data classification tasks.

\textbf{ECC} \cite{ecc} (Ensemble of Classifier Chains) employs a series of interconnected classifiers to enhance the precision of predictions, where each classifier uses its output as the input for the next classifier. This method is specifically designed for multi-label classification tasks and can effectively improve the overall performance of the model.

\textbf{RETAIN} \cite{retain} is an attention-based model tailored for sequence data analysis, adept at integrating temporal dynamics and specific features for accurate disease forecasting and disease. By dynamically capturing critical clinical events of a patient's history, RETAIN offers medication combinations.

\textbf{GAMENet} \cite{gamenet} is a medication recommendation that integrates the strengths of graph neural networks with memory networks and effectively discerns patterns and temporal sequences within medical data, thereby enhancing the precision of its predictions.

\textbf{SafeDrug} \cite{safedrug} leverages the combination of patients' health status and medication-related molecular knowledge. This approach, by reducing the impact of DDIs, can recommend safer medication combinations.

\textbf{MICRON} \cite{micron} focuses on customizing medication recommendation plans based on the dynamic changes in patient's health status. It does not produce new recommendations but updates medication combinations according to patients' new symptoms to enhance therapeutic effects while reducing potential side effects.

\textbf{COGNet} \cite{cognet} employs the Transformer architecture for medication recommendations, using a translation approach to infer medications from illnesses. It also features a copy mechanism to integrate beneficial medications from past prescriptions into new recommendations.

\textbf{MoleRec} \cite{molerec} delves into the importance of specific molecular substructures in medications. This approach enhances the precision of medication recommendations by leveraging finer molecular representations.

\textbf{CausalMed} \cite{causalmed} utilizes causal discovery based on patient status to identify primary and secondary diseases, thereby enhancing personalized patient representation.

\subsection{Evaluation Metrics}

We delve into the performance evaluation of our method using four principal metrics \cite{safedrug}: Jaccard index, DDI rate, F1 score, and PRAUC. These metrics provide comprehensive insights into the effectiveness and safety of our approach.

\textbf{Jaccard} (Jaccard Similarity Score) is employed to gauge the similarity between two sets. In medication recommendation, a higher Jaccard score indicates that the predicted prescription is more consistent with the actual medication regimen, indicating higher accuracy.
\begin{gather}
    Jaccard(t) = \frac{|\{i:\hat{m}_i=1\}|\cap|\{i:{m}_i=1\}|}{|\{i:\hat{m}_i=1\}|\cup |\{i:{m}_i=1\}|}, \\
    Jaccard = \frac{1}{N_h}\sum_{t=1}^{N_h} Jaccard(t),
\end{gather}
    where $\hat{m}_i$ represents the predicted outcome, $m_i$ represents the real label, and $N_h$ represents the total number of visits for patient $h$.

\textbf{DDI} (Drug-Drug Interaction Rate) measures the occurrence of interactions within the recommended combinations, a lower rate indicates higher safety of the medications.
\begin{equation}
    DDI = \frac{\sum_{i=1}^{N}\sum_{k,l\in \{j:\hat{m}_j(t)=1\}}1\{a^{ddi}_{kl}=1\}}{\sum_{i=1}^{N}\sum_{k,l\in \{ j:m_j(t)=1\}}1 },
\end{equation}
where $m(t)$ and $\hat{m}(t)$ denote the real and predicted multi-label at the visit $t$, $m_j(t)$ denotes the $j^{th}$ entry of $m(t)$, $a^{ddi}$ is the prior DDI relation matrix and $1$ is an indicator function which returns 1 when $a^{ddi}$ = 1, otherwise 0.

\textbf{F1} (F1-score) combines precision and recall, reflecting the model's ability to accurately identify correct medications while ensuring comprehensive coverage.
\begin{gather}
    Precision(t) = \frac{|\{i:\hat{m}_i=1 \}\cap \{i:m_i=1\}|}{|\{i:\hat{m}_i=1\}|},\\
    Recall(t) = \frac{|\{i:\hat{m}_i=1 \}\cap \{i:m_i=1\}|}{|\{i:m_i=1\}|},\\
    F1(t) = \frac{2}{\frac{1}{Precision(t)}+\frac{1}{Recall(t)}},\\
    F1 = \frac{1}{N_h}\sum_{i=1}^{N_h}F1(i).
\end{gather}

\textbf{PRAUC} (Precision-Recall Area Under Curve) assesses model performance across different recall levels, indicating the ability to maintain precision with increasing recall.
\begin{gather}
    PRAUC_t = \sum_{k=1}^{|M|} Precision_{k_t}\triangle Recall_{k_t},\\
    \triangle Recall_{k_t} = Recall_{k_t} - Recall_{{k-1}_t},
\end{gather}
where $|M|$ denotes the number of medications, $k$ is the rank in the sequence of the retrieved medications, and $Precision_{k}(t)$ represents the precision at cut-of $k$ in the ordered retrieval list and $\triangle Recall_{k_t}$ denotes the change in recall of a medication's ranking from $k-1$ to $k$. We averaged the PRAUC across all of the patient's visits.
\begin{equation}
    PRAUC = \frac{1}{N_h}\sum_{i=1}^{N_h}PRAUC(t).
\end{equation}

\textbf{Avg.\#Med} (Average number of medications) indicates the average medications per recommendation. A higher value suggests more complex combinations, potentially increasing the risk of adverse reactions, while a lower value indicates safer, more minimalistic treatment plans. This metric should be used as a reference only and not as a strict evaluative criterion.

\begin{equation}
    Avg.\#Med = \frac{1}{N_h}\sum_{i=1}^{N_h}|\hat{M}(i)|,
\end{equation}
where $|\hat{M}(i)|)$ denotes the number of predicted medications in visit $i$ of patient $h$.

\subsection{Implementation Details}

\subsubsection{Configuration and Parameter}
For all entity embeddings, the optimal dimension is set to 64. The GIN has only 1 layer and W-RGCN has 2 layers. The activation function of the MLP in this paper is the Sigmoid function, with a dropout rate of 0.5. In the bias correction module, the upper bound $\delta_1$ is set to 0.97, and the lower bound $\delta_2$ is set to 0.90. In the loss function, $\beta$ is set to 0.95, $kp$ is set to 0.05, and the acceptance rate $\gamma$ is set to 0.06. Training epoch $ep$=20, using the Adam optimizer with a learning rate of $lr$ = 0.0005 and a regularization factor $Reg$ = 0.05, and no batch is used in the training process.

\subsubsection{Sampling Approach}
Due to the limited availability of publicly accessible EHR data, we adopt bootstrapping sampling in this phase, following the approach recommended in \cite{safedrug}. This technique is particularly effective in scenarios where sample sizes are small, as detailed in \cite{testsample1} and \cite{testsample2}.

\subsubsection{Experimental Environment and Scalability}
\label{subsub:Experimental_Environmen}
The experimental environment employed in this study consists of an Ubuntu 22.04 operating system, a 12-core CPU with 64.3\% utilization on a single core, 3.4 GB of memory usage out of 30 GB available, and a 24 GB NVIDIA RTX 3090 GPU. The software dependencies include PyTorch version 2.0.0 and CUDA version 11.7.

\begin{figure}
    \centering
    \includegraphics[width=0.8\linewidth]{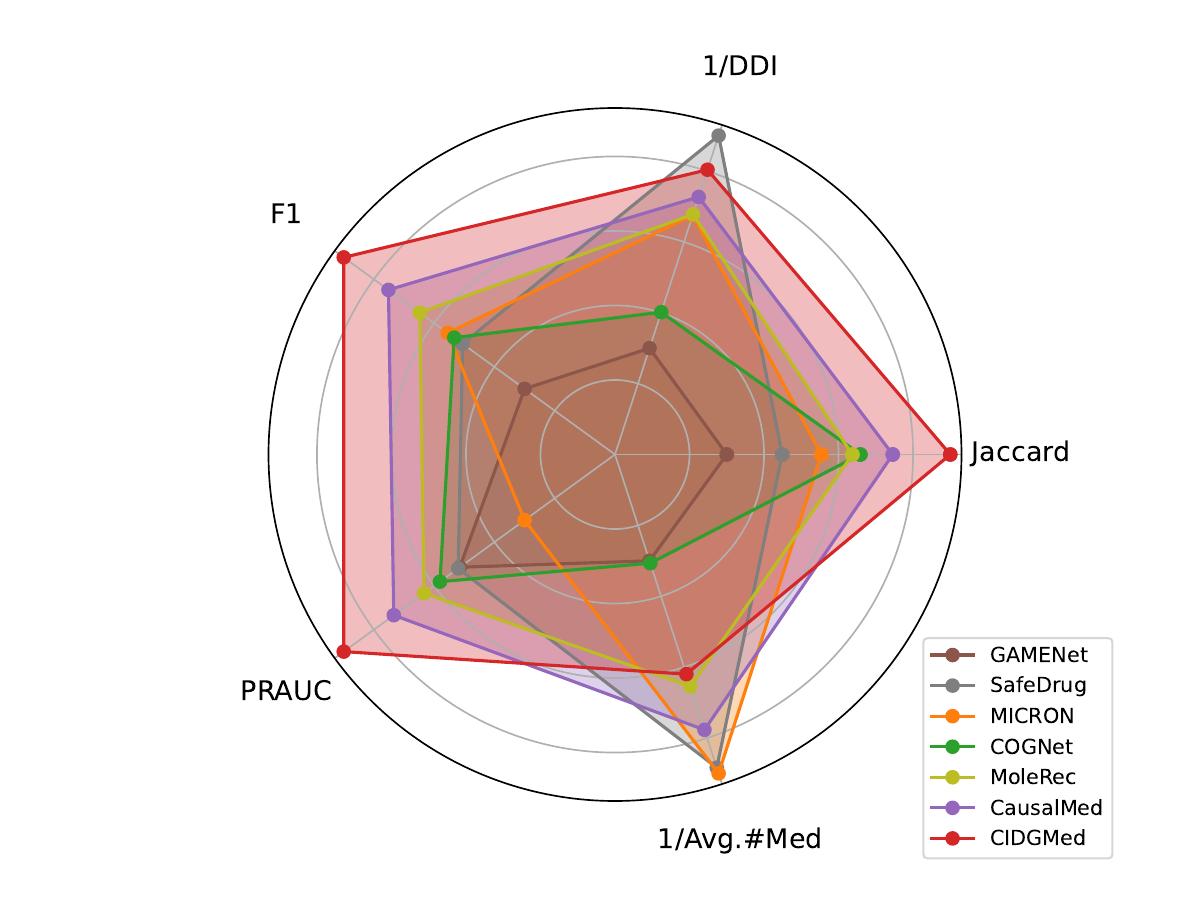}
    \caption{Comparison with recent outstanding works across all metrics in MIMIC-III.}
    \label{fig:comparison}
\end{figure}

\subsection{Performance Comparison (RQ1)}
In this section, we compare our model with baseline models, focusing on the accuracy, safety, and time efficiency of each.

\subsubsection{Effectiveness Analysis}

\begin{table*}
    \caption{The performance of each model on the MIMIC-III test set regarding accuracy and safety. The best and the runner-up results are highlighted in bold and underlined respectively under t-tests, at the level of 95\% confidence level.}
    \begin{tabular}{|*{1}{>{\centering\arraybackslash}p{2cm}}| *{5}{>{\centering\arraybackslash}p{2.5cm}}|}
    \toprule
    Model    & Jaccard$\uparrow$ & DDI$\downarrow$ & F1$\uparrow$ & PRAUC$\uparrow$ & Avg.\#Med \\
    \midrule
    LR          & 0.4924$\pm$0.0011    & 0.0830$\pm$0.0025    & 0.6490$\pm$0.0019    & 0.7548$\pm$0.0009    & 16.0489$\pm$0.0015 \\
    ECC         & 0.4856$\pm$0.0017    & 0.0817$\pm$0.0018    & 0.6438$\pm$0.0012    & 0.7590$\pm$0.0024    & 16.2578$\pm$0.0007 \\
    RETAIN      & 0.4871$\pm$0.0008    & 0.0879$\pm$0.0022    & 0.6473$\pm$0.0027    & 0.7600$\pm$0.0010    & 19.4222$\pm$0.0017 \\ 
    LEAP        & 0.4526$\pm$0.0007    & 0.0762$\pm$0.0015    & 0.6147$\pm$0.0021    & 0.6555$\pm$0.0014    & 18.6240$\pm$0.0019 \\
    GAMENet     & 0.4994$\pm$0.0013    & 0.0890$\pm$0.0006    & 0.6560$\pm$0.0016    & 0.7656$\pm$0.0023    & 27.7703$\pm$0.0018 \\
    SafeDrug    & 0.5154$\pm$0.0015    & \textbf{0.0655$\pm$0.0021}    & 0.6722$\pm$0.0011    & 0.7627$\pm$0.0008    & 19.4111$\pm$0.0022 \\
    MICRON      & 0.5219$\pm$0.0009    & 0.0727$\pm$0.0028    & 0.6761$\pm$0.0013    & 0.7489$\pm$0.0016    & 19.2505$\pm$0.0007 \\
    COGNet      & 0.5312$\pm$0.0018    & 0.0839$\pm$0.0005    & 0.6744$\pm$0.0023    & 0.7708$\pm$0.0014    & 27.6335$\pm$0.0011 \\
    MoleRec     & 0.5293$\pm$0.0020    & 0.0726$\pm$0.0024    & 0.6834$\pm$0.0008    & 0.7746$\pm$0.0012    & 22.0125$\pm$0.0006 \\
    CausalMed   & \underline{0.5389$\pm$0.0011}   & 0.0709$\pm$0.0007  & \underline{0.6916$\pm$0.0022}   & \underline{0.7826$\pm$0.0010}   & 20.5419$\pm$0.0028 \\
    \midrule
    \textbf{CIDGMed}   & \textbf{0.5526$\pm$0.0016}   & \underline{0.0684$\pm$0.0014}  & \textbf{0.7033$\pm$0.0009}   & \textbf{0.7955$\pm$0.0026}    & 22.4693$\pm$0.0013 \\
    \bottomrule
    \end{tabular}
    \label{tab:mimic3}
\end{table*}

\begin{table*}
    \caption{The performance of each model on the MIMIC-IV test set regarding accuracy and safety. The best and the runner-up results are highlighted in bold and underlined respectively under t-tests, at the level of 95\% confidence level.}
    \begin{tabular}{|*{1}{>{\centering\arraybackslash}p{2cm}}| *{5}{>{\centering\arraybackslash}p{2.5cm}}|}
    \toprule
    Model    & Jaccard$\uparrow$ & DDI$\downarrow$ & F1$\uparrow$ & PRAUC$\uparrow$ & Avg.\#Med \\
    \midrule
    LR          & 0.4569$\pm$0.0011    & 0.0783$\pm$0.0013    & 0.6064$\pm$0.0016    & 0.6613$\pm$0.0019    & 8.5746$\pm$0.0012 \\
    ECC         & 0.4327$\pm$0.0014    & 0.0764$\pm$0.0011    & 0.6129$\pm$0.0018    & 0.6530$\pm$0.0017    & 8.7934$\pm$0.0010 \\
    RETAIN      & 0.4234$\pm$0.0017    & 0.0936$\pm$0.0015    & 0.5785$\pm$0.0013    & 0.6801$\pm$0.0012    & 10.9576$\pm$0.0018 \\ 
    LEAP        & 0.4254$\pm$0.0013    & 0.0688$\pm$0.0011    & 0.5794$\pm$0.0017    & 0.6059$\pm$0.0018    & 11.3606$\pm$0.0012 \\
    GAMENet     & 0.4565$\pm$0.0018    & 0.0898$\pm$0.0012	& 0.6103$\pm$0.0019	& 0.6829$\pm$0.0017    & 18.5895$\pm$0.0015 \\
    SafeDrug    & 0.4487$\pm$0.0012    & \textbf{0.0604$\pm$0.0010}    & 0.6014$\pm$0.0017    & 0.6948$\pm$0.0018    & 13.6943$\pm$0.0014 \\
    MICRON      & 0.4640$\pm$0.0017    & 0.0691$\pm$0.0015    & 0.6167$\pm$0.0016    & 0.6919$\pm$0.0014    & 12.7701$\pm$0.0011 \\
    COGNet      & 0.4775$\pm$0.0014    & 0.0911$\pm$0.0013	& 0.6233$\pm$0.0019	& 0.6524$\pm$0.0018    & 18.7235$\pm$0.0015 \\
    MoleRec     & 0.4744$\pm$0.0013    & 0.0722$\pm$0.0014	& 0.6262$\pm$0.0018	& 0.7124$\pm$0.0017    & 13.4806$\pm$0.0015  \\
    CausalMed   & \underline{0.4899$\pm$0.0014}	& 0.0677$\pm$0.0017  & \underline{0.6412$\pm$0.0013} & \underline{0.7338$\pm$0.0019}  & 14.4295$\pm$0.0012 \\
    \midrule
    \textbf{CIDGMed}   & \textbf{0.5019$\pm$0.0012}	 & \underline{0.0644$\pm$0.0016}	 & \textbf{0.6524$\pm$0.0014}	   & \textbf{0.7449$\pm$0.0018} & 18.4956$\pm$0.0015\\
    \bottomrule
    \end{tabular}
    \label{tab:mimic4}
\end{table*}

Table ~\ref{tab:mimic3} and \ref{tab:mimic4} details the results of the effectiveness comparison. Utilizing traditional machine learning approaches, LR and ECC demonstrate modest accuracy and, despite prescribing fewer medications, still encounter a higher DDI rate. In contrast, sequence-based models like LEAP, which employ advanced deep learning techniques, fail to surpass these traditional methods in effectiveness, highlighting the potential limitations of generative models in this domain. Furthermore, RETAIN, which introduces sequence models in medication recommendation, neglects the interactions between medications, consequently leading to an increased DDI rate.

We conduct a comprehensive comparison and discussion of several recent and high-performing baselines, as shown in Figure~\ref{fig:comparison}. It is important to note that there are substantial differences in the metrics across different models. For ease of presentation, these metrics undergo standard normalization. Among the five indicators, the Jaccard index, F1 score, and PRAUC are used for accuracy testing, while DDI and Avg.\#Med are used for safety testing. Considering that lower values of DDI and the Average Number of medications indicate improved safety, their reciprocals are represented in Figure~\ref{fig:comparison}.

First, GAMENet, COGNet, and MICRON do not perform well in terms of accuracy and safety. Although COGNet shows a decent Jaccard index, its performance on other accuracy metrics is poor. MICRON performs reasonably well on the Avg.\#Med metric, but its DDI metric is unsatisfactory. This is because, despite considering patient history, these models do not fully explore the fine-grained relationships between diseases/procedures and medications, resulting in some errors in the recommendation outcomes and affecting the models' performance.

Secondly, SafeDrug and MoleRec conduct detailed analyses of fine-grained molecular-level data of medications. While their accuracy needs improvement, their safety is guaranteed. This proves that fine-grained molecular structures of medications can help models avoid the adverse effects of medication interactions during recommendations. However, they only use a single-granularity approach and do not consider or explore the direct impact of the medication as a whole (coarse-granularity) on the disease or procedure.
CausalMed is a patient-centred representation that emphasizes sequencing between diseases through causal discovery with a significant improvement in accuracy, but fails to take into account medication-patient associations at the molecular level of action.
Consequently, their effectiveness is not as good as the method proposed in this paper, CIDGMed.

Therefore, our CIDGMed achieves excellent results in both accuracy and safety, further demonstrating the effectiveness of our approach by exploring causal relationships and combining dual-granularity information.

\subsubsection{Time Efficiency Analysis}
\begin{table}
  \caption{The performance of recent excellent models in training and inference efficiency.}
  \begin{tabular*}{\tblwidth}{@{} CCCCC@{} }
   \toprule
    Model & \makecell{Convergence\\ Epoch}  & \makecell{Training Time\\/Epoch(s)} & \makecell{Total Training\\Time(s)} & \makecell{Inference\\ Time(s)} \\
   \midrule
    GAMENet & 39 & 45.31 & 1767.09 & 19.27\\
    SafeDrug & 54 & 38.32 & 2069.28 & 20.15\\
    MICRON & 40 & 17.48 & 699.20 & 14.48\\ 
    COGNet & 103 & 38.85 & 4001.55 & 142.91\\
    MoleRec & 25 & 249.32 &	6233.00 & 32.10\\
    CausalMed & 33 & 164.77 & 5437.41 & 18.29  \\
    \midrule
    CIDGMed & 10 & 329.41 & 3294.10 & 21.85\\
   \bottomrule
  \end{tabular*}
  \label{tab:time}
\end{table}

As shown in Table~\ref{tab:time}, this paper compares and studies the time efficiency of CIDGMed in terms of average training time per epoch, convergence epoch, total training time, and inference time against representative recommendation algorithms. To ensure fairness, all experiments are conducted in the experimental environment mentioned in~\ref{subsub:Experimental_Environmen}. From the table~\ref{tab:time} and Figure~\ref{fig:comparison}, we observe that CIDGMed achieves a better balance between time efficiency, recommendation accuracy, and safety.

Additionally, we present the time complexity of CIDGMed in detail. Specifically,
In the relationship mining module, the main task is to utilize causal discovery to unearth the relationships between various entities and quantify these relationships through causal inference, with a time complexity of \(O(n \cdot {dim}^2)\), where \(n\) represents the number of samples and \({dim}\) represents the dimension of the features.

Next, the representation learning stage employs attention mechanisms, GIN, and W-W-RGCN for computation, along with GRU and MLP. The time complexity of this stage is primarily influenced by the length of the input sequence (\(n\)) and the dimension of the model (\({dim}\)). The time complexity of the attention mechanism is \(O(n^2 \cdot {dim})\), while that of GIN and W-RGCN is \(O(n^2)\). Additionally, the computational cost of integrating embeddings through adaptive classifiers and attention mechanisms is \(O((n^2 + 1) \cdot {dim})\). The time complexity of GRU and MLP is related to the time steps \(T\) of each GRU unit, expressed as \(O(T \cdot {dim}^2 + {dim}^2)\). Therefore, the overall time complexity of the representation learning stage can be expressed as \(O(T \cdot {dim}^2 + {dim}^2 + n^2 \cdot {dim} + {dim}+ n^2)\).

Finally, in the bias correction stage, the data originates from the causal inference. The main task of this stage is to evaluate and adjust probability values, with a time complexity of \(O(n)\).
Ultimately, the total complexity of the entire model is: 
\begin{equation}
   O((n+T+1){dim}^2 +(n^2+1){dim}+n^2+n).
\end{equation}

In analyzing the time consumption of models, we observe that models with higher time costs often employ complex network structures of up to 3 to 5 layers, especially in the application of GNNs. In contrast, our design graph network construction method requires fewer layers, thereby significantly reducing the time overhead. Moreover, the bias correction mechanism we propose utilizes a method based on statistical principles and rules, which greatly alleviates the computational burden.

When assessing through both individual metrics and an overall view, CIDGMed significantly outshines other baseline models. Utilizes the best current model CausalMed as a comparative benchmark, CIDGMed notably lowers safety risks by reducing the DDI rate by 3.65\%, boosts accuracy metric Jaccard by 2.54\%, and also cuts down on training time by 39.42\%, showcasing its efficiency and effectiveness in medication recommendation.

\subsection{Ablation Study (RQ2\&RQ3\&RQ4)}
To assess the contribution of different components of the proposed model CIDGMed, we test their importance by removing key modules.

\begin{table*}
\caption{The performance of each ablation model on the test set regarding accuracy and safety. The best and the runner-up results are highlighted in bold and underlined respectively under t-tests, at the level of 95\% confidence level.}
    \begin{tabular}{|*{1}{>{\centering\arraybackslash}p{3.3cm}}| 
    *{4}{>{\centering\arraybackslash}p{1.2cm}}|
    *{4}{>{\centering\arraybackslash}p{1.2cm}}|}
    \toprule
    \multirow{2}{*}{Model}
    & \multicolumn{4}{c|}{MIMIC-III} & \multicolumn{4}{c|}{MIMIC-IV} \\ 
    \cmidrule(lr){2-5} \cmidrule(lr){6-9}
    & Jaccard$\uparrow$ & DDI$\downarrow$ & F1$\uparrow$ & PRAUC$\uparrow$  & Jaccard$\uparrow$ & DDI$\downarrow$ & F1$\uparrow$ & PRAUC$\uparrow$  \\
    \midrule
    CIDGMed $w/o$ C        & 0.5484	 & 0.0689	& 0.6954	& 0.7902 & \underline{0.4874}   & 0.0673    & 0.6382	& 0.7274	 \\
    CIDGMed $w/o$ F        & \underline{0.5494}	 & 0.0726	& \underline{0.6992}	& \underline{0.7905}	  & 0.4863    & 0.0701    & \underline{0.6389}	& \underline{0.7318}	 \\
    CIDGMed $w/o$ C+F        & 0.5477	& 0.0724	& 0.6924	& 0.7895	 & 0.4856	& 0.0702	& 0.6366	& 0.7254  \\
    CIDGMed $w/o$ BC        & 0.5357	& \textbf{0.0672}    & 0.6888	& 0.7815	 & 0.4798	& \underline{0.0637}	& 0.6307	& 0.7244 \\
    CIDGMed $w/o$ C+F+BC      & 0.5336	& 0.0721	& 0.6875	& 0.7781	  & 0.4836    & 0.0649    & 0.6346	& 0.7209	 \\
    \midrule
    \textbf{CIDGMed}   & \textbf{0.5526}   & \underline{0.0684}  & \textbf{0.7033}   & \textbf{0.7955}   & \textbf{0.5019}	 & \textbf{0.0644}	 & \textbf{0.6524}	  & \textbf{0.7449} \\
    \bottomrule
    \end{tabular}
    \label{tab:ablation}
\end{table*}

\textbf{CIDGMed $w/o$ C}: removes the coarse-grained relationship learning based on causal inference in the representation learning module. The relationships between diseases/procedures and medications use traditional co-occurrence relationships, meaning there is an edge between each pair of diseases/procedures and medications, with edge weights derived from co-occurrence rates.

\textbf{CIDGMed $w/o$ F}: removes fine-grained relationships from the molecular structure level in the representation learning module. This eliminates structural connections between medications. Medication representations are obtained through random initialization.

\textbf{CIDGMed $w/o$ C+F}: simultaneously removes the representation learning methods for both coarse-grained and fine-grained relationships based on causal inference. Instead, medication representations are obtained purely through random initialization while constructing the graph network using co-occurrence rates.

\textbf{CIDGMed $w/o$ BC}: removes the bias correction module. The model does not perform bias correction based on causal relationships and uses the initial probabilities obtained from representation learning to recommend medication combinations.

\textbf{CIDGMed $w/o$ C+F+BC}: removes most of the innovations from this paper. The model constructs graphs and recommends medications solely based on co-occurrence rates, without considering causal relationships or performing bias correction.

The results of the ablation study are shown in Table~\ref{tab:ablation}. Specifically, \textbf{CIDGMed $w/o$ C} significantly decreases recommendation accuracy, indicating that the coarse-grained method based on causal inference can uncover the true causal relationships between medication entities, playing an essential role in improving overall accuracy. \textbf{CIDGMed $w/o$ F} significantly decreases safety while slightly reducing recommendation accuracy, indicating that the fine-grained relationships based on molecular structure provide a crucial representation foundation for the safety relationships between medications and play a significant role in reducing safety hazards. The results of \textbf{CIDGMed $w/o$ C+F} show that removing both coarse-grained and fine-grained representations significantly decreases performance. Thus, combining these representations compensates for deficiencies in safety and accuracy, leading to better and more balanced outcomes.

The results of \textbf{CIDGMed $w/o$ BC} show a substantial decrease in accuracy, clearly indicating that the bias correction module significantly improves accuracy. This improvement is due to the module combining the causal relationships between diseases/procedures and medications, performing post-intervention processing during the recommendation phase to correct biases in the model learning process, ultimately enhancing accuracy. Additionally, safety shows a slight improvement.

The result of \textbf{CIDGMed $w/o$ C+F+BC} shows that all metrics decline, clearly indicating a synergistic effect among our proposed innovations, and combining them results in significantly better improvements.

In summary, the results of multiple ablation experiments demonstrate the rationality and effectiveness of our modules, indicating that the design of the causal-based module is indispensable as it significantly enhances the model's accuracy and safety. Additionally, the dual-grained framework ensures more accurate and safer medication combinations, with the coarse-grained approach better serving model accuracy, while the fine-grained approach enhances the safety of medication recommendations. We also observe that in the field of medication recommendation, safety and accuracy often constrain each other, and achieving optimal results typically involves balancing both aspects effectively.

\subsection{Parameter Sensitivity}

\begin{figure*}
    \centering
    \begin{subfigure}{\linewidth}
        \includegraphics[width=\textwidth]{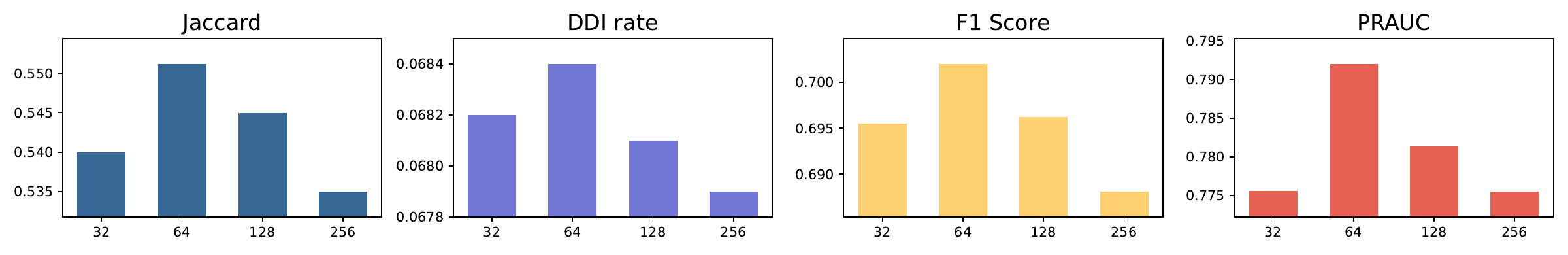}
        \vspace{-0.7cm}
        \caption{The impact of embedded dimensions on results.}
        \label{fig:para_embedding}
    \end{subfigure}

    \begin{subfigure}{\linewidth}
        \includegraphics[width=\textwidth]{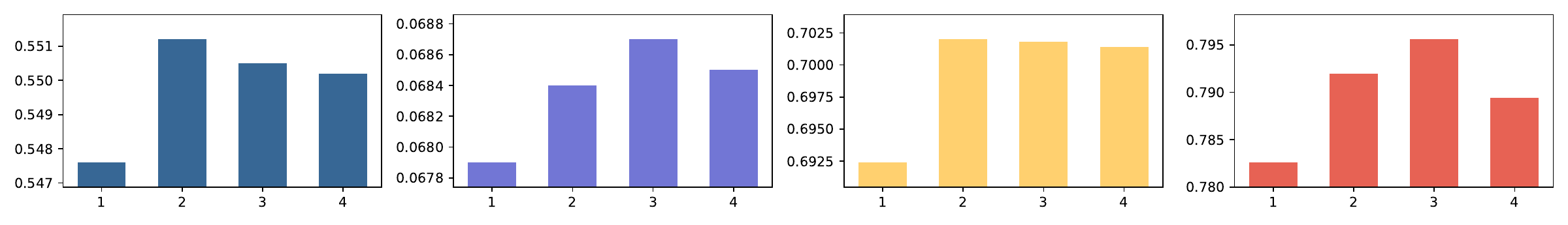}
        \vspace{-0.7cm}
        \caption{The impact of the number of W-RGCN layers on results.}
        \label{fig:para_rgcnLayer}
    \end{subfigure}
    
    \begin{subfigure}{\linewidth}
        \includegraphics[width=\textwidth]{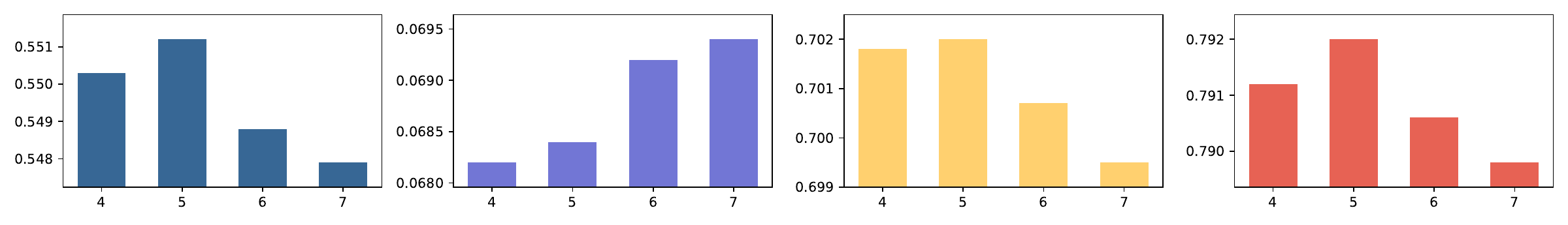}
        \vspace{-0.7cm}
        \caption{The impact of the number of relationship types in W-RGCN on results.}
        \label{fig:para_relationshipTypeNum}
    \end{subfigure}

    \begin{subfigure}{\linewidth}
        \includegraphics[width=\textwidth]{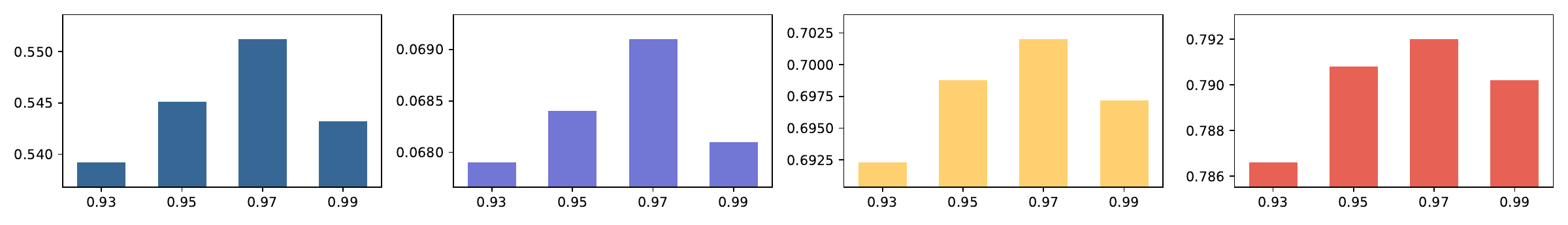}
        \vspace{-0.7cm}
        \caption{The impact of upper bound on results.}
        \label{fig:para_upper}
    \end{subfigure}

    \begin{subfigure}{\linewidth}
        \includegraphics[width=\textwidth]{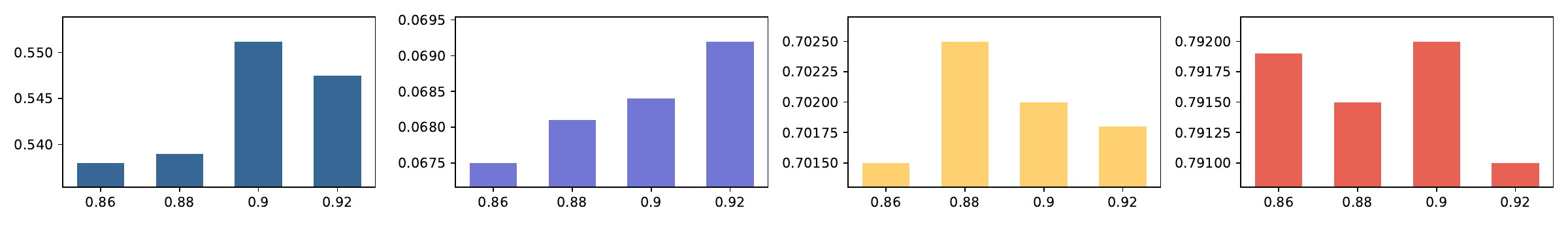}
        \vspace{-0.7cm}
        \caption{The impact of lower bound on results.}
        \label{fig:para_lower}
    \end{subfigure}

    \vspace{-0.2cm}
    \caption{Experiments on parameter sensitivity on MIMIC-III dataset.}
    \label{fig:para_sensitive}
\end{figure*}

To thoroughly explore the impact of model parameters on performance, we conduct a sensitivity analysis on five key parameters using the MIMIC-III dataset. The results, shown in Figure~\ref{fig:para_sensitive}, demonstrate consistent trends in both MIMIC-III and MIMIC-IV validations. However, due to space limitations, we only present the detailed validation results for MIMIC-III.

Figure~\ref{fig:para_embedding} shows the model's performance under different embedding dimensions. When the embedding dimension is set to 64, the model's accuracy significantly surpasses other settings, and its safety performance is nearly identical to that at 128. As mentioned earlier, increasing the embedding dimension raises the model's time cost quadratically. Thus, considering accuracy, safety, and time cost, we select 64 as the optimal embedding dimension.

Secondly, the paper focuses on two core parameters when constructing the W-RGCN graph network: the number of layers and the number of node categories. The impact of the W-RGCN layers is illustrated in Figure \ref{fig:para_rgcnLayer}, showing that the model performs extremely poorly in accuracy when the number of layers is 1. While the accuracy performance does not differ much between 2, 3, and 4 layers, lower numbers of layers bring better safety. This is because an increase in layers can lead to overfitting, causing the model to overly pursue accuracy in combinations while neglecting medication safety, potentially leading to a slightly higher DDI index. Additionally, an increase in layers significantly raises the time cost. Hence, after comprehensive consideration, we chose 2 layers as the best setting.

The results regarding the number of relationships in W-RGCN categories are shown in Figure \ref{fig:para_relationshipTypeNum}. Finer division leads to fewer categories, and many different relationships are classified into the same category, making it difficult to express the differences between them. On the other hand, if there are too few categories, each category might lack enough samples, hindering the model's ability to be adequately trained and capture a generalized representation at that level. Therefore, dividing into 5 categories was chosen as the optimal setting.

Regarding the setting of upper and lower bounds, it is crucial for the function of the probability adjustment module. It is worth mentioning that, in this experiment, both the upper and lower bounds were tested by incrementing from 0.5 to 1 in steps of 0.01. Ultimately, the four most representative values were selected for presentation. Figures \ref{fig:para_lower} and \ref{fig:para_upper} demonstrate that when the causal effect of a medication on a specific disease/procedure is below 0.90, such medication can be considered to have a low association with the disease, implying that the data correlation between the medication and disease is not directly caused by their causal relationship but may be due to co-occurrence phenomena caused by other entities slightly related to them. Conversely, when the causal effect is above 0.97, we regard it as having a direct causal relationship.

\subsection{Case Study (RQ4\&RQ5)}
\label{subsub:case study}

\begin{figure}
    \centering
    \includegraphics[width=1\linewidth]{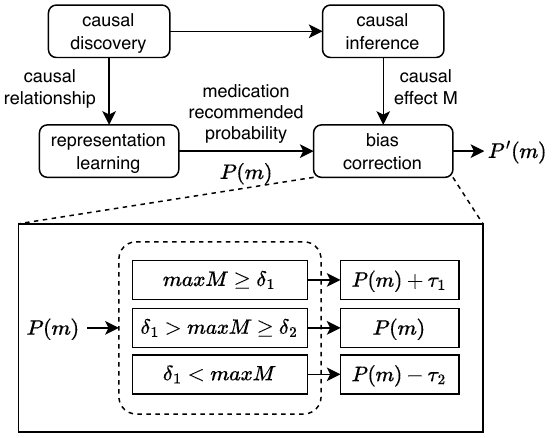}
    \caption{Detailed workflow for bias correction.}
    \label{fig:case3}
\end{figure}

\begin{figure*}
    \centering
    \includegraphics[width=0.95\textwidth]{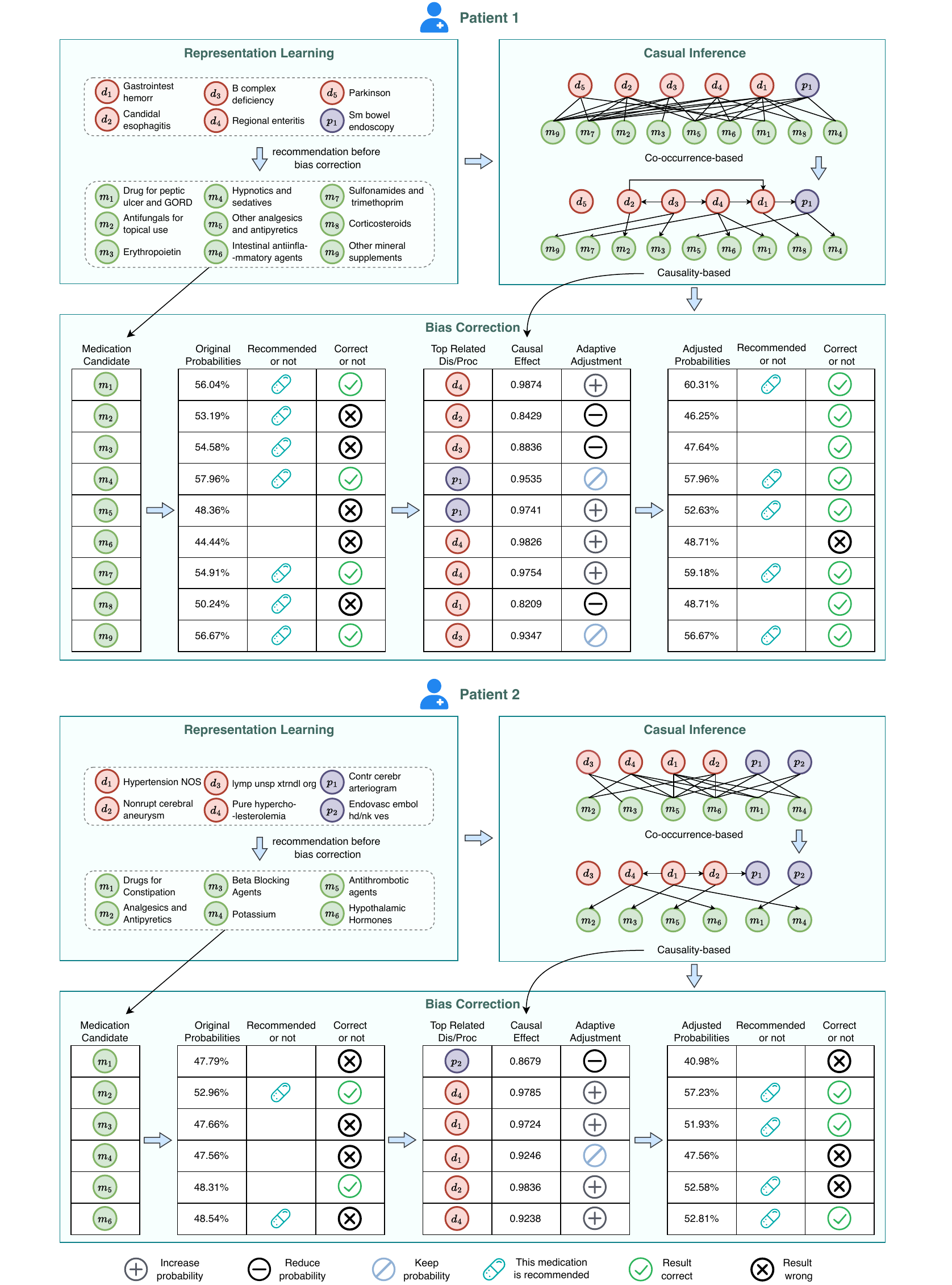}
    \caption{A detailed explanation of the process in this paper, using the real cases of two patients as examples.}
    \label{fig:case2}
\end{figure*}

Figure \ref{fig:case3} illustrates the workflow of this paper, especially the explanation of the bias correction process. To further elucidate the effectiveness and rationality of the model based on causal inference, we randomly selected some real cases in the MIMIC-III dataset. 

As shown in figure \ref{fig:case2}, we randomly select two patients ($Patient 1$ and $Patient 2$) from our dataset to demonstrate the recommendation process based on the method proposed in this paper.


\subsubsection{Causal Inference}
We use traditional co-occurrence analysis methods and causal inference techniques, respectively, to explore the associations between these diseases and treatments. The casual inference part in figure \ref{fig:case2} shows the relationships between diseases/ procedures and candidate medications. 
Taking patient 1 as an example, in the traditional co-occurrence-based relationship graph, almost every disease/procedure has a direct connection with medications, resulting in an overly complex and ambiguous network. Thus, the co-occurrence method cannot effectively capture the true underlying relationships between different entities, severely affecting the accuracy and safety of the recommendations and making it difficult to explain the reasons for the recommendations. In contrast, causal inference significantly clarifies the relationships between medications and diseases/procedures. For instance, in the co-occurrence relationships, $m_9$ is connected to $d_2$, $d_3$, $d_1$, and $p_1$, but causal inference reveals that $m_9$ (other mineral supplements) is only related to $d_3$ (B complex deficiency).

\subsubsection{Bias Correction}

\chen{We use the case of two patients to illustrate how bias correction is performed in this study. The detailed process is illustrated in Figure~\ref{fig:case2}. Specifically, by reviewing their current disease conditions and procedure records, we first illustrate how the proposed method originally recommends medications (before bias correction) for each patient. Next, the bias correction module then utilizes the causal matrix to refine and adjust these original recommendations. This step-by-step process provides a clear demonstration of how the system works, from the original recommendation to the correction phase, ultimately enhancing the accuracy of the medication recommendations.}

\chen{Through an in-depth analysis of the recommendation process using the bias correction module for $Patient 1$ and $Patient 2$, we summarize it into the following three scenarios.}

\chen{Scenario 1: causal effect greater than the upper bound of 0.97. These medications have a highly relevant causal effect with the disease or procedure (greater than 0.97), indicating a strong connection to the patient's condition. As a result, they are considered more suitable for the patient, and their recommendation probability is increased accordingly. For example: medications $m_1$, $m_5$, and $m_7$ in $Patient 1$, as well as $m_2$, $m_3$, and $m_6$ in $Patient 2$.}

\chen{Scenario 2: causal effect less than the lower bound of 0.90. These medications exhibit no significant causal effect with the disease or procedure (less than 0.90), making them less likely to be relevant to the patient's condition. As a result, their recommendation probability is reduced to avoid unnecessary or ineffective treatments. For example: medications $m_2$, $m_3$, and $m_8$ in $Patient 1$.}

\chen{Scenario 3: causal effect between 0.90 and 0.97. These medications have a causal effect with the disease or procedure between 0.90 and 0.97, so no adjustment in their recommendation probability is needed, and the recommendation remains unchanged. For example: medications $m_4$ and $m_9$ in $Patient 1$, and $m_4$ in $Patient 2$.}

\chen{In summary, bias correction of medications based on the causal matrix between diseases/procedures and medications can significantly improve the performance of medication recommendations. Additionally, the in-depth analysis presented above enhances the interpretability of the proposed method, further validating its effectiveness.}

\subsection{Error Analysis and Limitation}
\chen{We perform an error analysis of the model’s recommendation results to better understand its limitations and identify areas for improvement. The prescription errors observed can be classified into two categories:}

\chen{The first type of error is \textbf{False Negative (FN)}, where the model fails to recommend necessary medications, leading to omissions. This occurs mainly due to the infrequent appearance of certain medications in the records, which is also known as data sparsity in recommendation systems, making it difficult for the model to learn their relationships with associated diseases or procedures.}

\chen{The second type of error is \textbf{False Positive (FP)}, where the model recommends medications that should not have been prescribed. This mainly occurs because certain disease or procedure and medication combinations are imbalanced in the training data, causing the model to favour recommending frequently occurring medications, leading to over-recommendation.}

\begin{table}
    \centering
    \caption{The statistical analysis of the error in our method across both datasets.}
    \begin{tabular}{c|c|c}
        \toprule
        Dataset     & FN rate   & FP rate \\
        \midrule
        MIMIC-III   & 0.2392 $\pm$ 0.0050   &   0.0652 $\pm$ 0.0006 \\
        MIMIC-IV    & 0.2489 $\pm$ 0.0031   &   0.0712 $\pm$ 0.0008\\
        \bottomrule
    \end{tabular}
    \label{tab:error}
\end{table}

\chen{For these two cases, we conduct a statistical analysis of error, as shown in table \ref{tab:error}. We find that the FN value is slightly higher than the FP not only due to the data sparsity but also because medication recommendation involves suggesting a set of medications. If any medication in the set is not recommended, it is considered an FN. This differs from other domains, where only a single item is recommended, resulting in a higher FN value in our case. Although the FP is higher than the FN, it remains within a reasonable range, as both the accuracy and safety of our model are greatly improved compared to other baselines. In the future, we aim to further resolve the issue of over-recommendation.}

\section{Conclusion and Future Work}

This paper introduces our developed medication recommendation system, CIDGMed. By applying causal inference, CIDGMed uncovers the causal relationships between diseases/procedures and medications at both the coarse-grained and fine-grained levels, achieving dual-granularity feature fusion. Furthermore, CIDGMed employs a post-processing intervention method, bias correction, during model recommendation. Through a series of rigorous experiments on publicly available clinical datasets, we significantly improve the model's accuracy and safety. Additionally, the case study analysis further validates the rationality of CIDGMed. In summary, the results fully demonstrate the superior performance of our method in terms of accuracy, safety, time efficiency, and rationality.

However, it has not yet been evaluated in real medical environments due to ethical considerations and data protection concerns. In the future, we aim to deploy this method in practical settings and engage with domain experts to refine our approach. Their feedback will be crucial for optimizing our framework to meet the specific needs of real-world medical applications, thereby enhancing its practical value and effectiveness.

\section*{Declaration of Competing Interest}
The authors declare that they have no known competing financial interests or personal relationships that could have appeared to influence the work reported in this paper.

\printcredits
\section*{Acknowledgments}
\subsection*{Funding}
This work was supported by the Innovation Capability Improvement Plan Project of Hebei Province (No. 22567637H), the S\&T Program of Hebei(No. 236Z0302G), and HeBei Natural Science Foundation under Grant (No.G2021203010 \& No.F2021203038).

\bibliographystyle{cas-model2-names}

\bibliography{CIDGMed}

\end{document}